\documentclass[useAMS,usenatbib]{mnras}
\usepackage{graphicx}
\usepackage{epstopdf,color,verbatim}
\usepackage{threeparttable}
\usepackage[]{txfonts}

\def\simless{\mathbin{\lower 3pt\hbox
{$\rlap{\raise 5pt\hbox{$\char'074$}}\mathchar"7218$}}}   
\def\simmore{\mathbin{\lower 3pt\hbox
{$\rlap{\raise 5pt\hbox{$\char'076$}}\mathchar"7218$}}}   
\newcommand{\frb}{FRB 121102 }

\newcommand{\ompt}{\omega_{\rm p}t}

\usepackage{epsfig} 
\usepackage{epstopdf}

\title[Synchrotron maser from relativistic shocks in FRBs]
{The synchrotron maser emission from relativistic shocks in Fast Radio Bursts: 1D PIC simulations of cold pair plasmas}

\author[I. Plotnikov \& L. Sironi ]
{Illya Plotnikov$^{1,2}$ \thanks{E-mail: illyap@astro.princeton.edu}\ and Lorenzo Sironi$^3$  \thanks{E-mail: lsironi@astro.columbia.edu} \\
$^1$ Department of Astrophysical Sciences, Princeton University, 4 Ivy Ln., Princeton, NJ 08544, USA \\
$^2$ Universit\'e de Toulouse, UPS-OMP, IRAP, 9 av. Colonel Roche, BP 44346, F-31028 Toulouse Cedex 4, France \\
$^3$ Department of Astronomy, Columbia University, 550 W 120th St., New York, NY 10027, USA}

\begin{document}
\date{Received / Accepted}
\pagerange{\pageref{firstpage}--\pageref{lastpage}} \pubyear{2018}

\maketitle

\label{firstpage}

\begin{abstract}
The emission process of Fast Radio Bursts (FRBs) remains unknown. We investigate whether  the synchrotron maser emission from relativistic shocks in a magnetar wind can explain the observed FRB properties. We perform particle-in-cell (PIC) simulations of perpendicular shocks in cold pair plasmas, checking our results for consistency among three PIC codes. We confirm that a linearly polarized X-mode wave is self-consistently generated by the shock and propagates back upstream as a precursor wave. We find that at magnetizations $\sigma\gtrsim 1$ (i.e., ratio of Poynting flux to particle energy flux of the pre-shock flow) the shock converts a fraction $f_\xi' \approx 7 \times 10^{-4}/\sigma^2$ of the total incoming energy into the precursor wave, as measured in the shock frame. The wave spectrum is narrow-band (fractional width $\lesssim 1-3$), with apparent but not dominant line-like features as many resonances concurrently contribute. The peak frequency in the pre-shock (observer) frame is $\omega^{\prime \prime}_{\rm peak} \approx 3 \gamma_{\rm s | u} \omega_{\rm p}$, where $\gamma_{\rm s|u}$ is the shock Lorentz factor in the upstream frame and $\omega_{\rm p}$ the plasma frequency. 
At $\sigma\gtrsim1$, where our estimated $\omega''_{\rm peak}$ differs from previous works,  the shock structure presents two solitons separated by a cavity, and the peak frequency corresponds to an eigenmode of the cavity. 
Our results provide physically-grounded inputs for FRB emission models within the magnetar scenario.
\end{abstract} 


\section{Introduction}
Synchrotron masers are known to produce strong decametric radio emission in the Jovian magnetosphere and kilometric emission in the terrestrial magnetosphere (Auroral Kilometric Emission).
They are driven by a ``population inversion'' of energetic electrons gyrating in an intense magnetic field. The driver of the emission is either a loss-cone or ring-like electron  distribution function \citep{melrose_17, treumann_06}. Such a population inversion occurs also in strongly magnetized relativistic perpendicular shocks, where a coherent cold-ring distribution of particles is self-consistently produced as part of the shock evolution \citep{gallant_92, amato_arons_06}. This distribution is unstable to the synchrotron maser instability \citep{hoshino_91} that causes the emission of a train of high amplitude semi-coherent electromagnetic waves propagating from the shock front into the unshocked (upstream) medium \citep{gallant_92, hoshino_92}. The possible importance of the synchrotron maser in astrophysical sources was anticipated long ago \citep{sazonov_73}. Yet, to our present knowledge, there is no firm demonstration that  shock-powered synchrotron maser emission is dominant in any non-heliospheric astrophysical environment, even though a few potential scenarios have been proposed \citep[e.g.,][]{sagiv_02}. 

 Recently, however, the discovery of Fast Radio Bursts \citep[FRBs; ][]{lorimer_07, keane_12, thornton_13, spitler_14, marcote_2017} has revived the interest in this mechanism. These events are bright ($\sim 1\,$Jy) pulses of millisecond duration detected in the $\sim$~GHz band. Their extremely high brightness temperature, $T_{\rm B} \sim 10^{37}$K, requires a coherent emission mechanism \citep{katz_16, popov_18}. Young magnetars have emerged as one of the most likely  progenitors of FRBs, at least for the repeating class \citep[e.g., ][]{popov_13, murase_16,lyutikov_17, metzger_17, kashiyama_17, long_peer_18,margalit_18a, margalit_18b}. 
 Magnetars can naturally explain the short FRB durations, large energy requirements, and ordered magnetic fields needed for coherent emission. Thus far, most works on the FRB emission mechanism are based on mere considerations of energetics and timescales, in which it is assumed that a fraction of the free energy of the system is radiated away at GHz frequencies by coherent charge ``bunches'' via, e.g., curvature or synchrotron maser processes. In the  case of curvature radiation, the emission is postulated to be a product of magnetic reconnection close to the magnetar surface \citep[e.g.,][]{lyutikov_02, kumar_17, lu_kumar_18, ghisellini_18, katz_18}. In the case of the synchrotron maser, the emission is thought to occur at  relativistic shocks propagating in the magnetar wind or nebula \citep{lyubarsky_14, murase_16, belo_17} or inside the ultra-relativistic shell ejected from the central compact object \citep{waxman_17}.  Yet, in either case the conditions for coherent emission and the very existence of  charge bunches with the required properties are often postulated \emph{ad hoc}, resulting in models with little predictive power.

 The purpose of this work is to demonstrate from first principles that the synchrotron maser at relativistic shocks in the magnetar wind can naturally explain the observed FRB properties. By means of particle-in-cell (PIC) simulations, we investigate how the efficiency and spectrum of the electromagnetic wave emitted by the shock into the pre-shock medium (which we shall call ``precursor wave'') depend on the  physical conditions in the magnetar wind. In this work, the first of a series, we present results from one-dimensional (1D) simulations (more precisely, 1D3V, i.e., we employ one spatial dimension, but all three components of velocities and electromagnetic fields are retained), while multi-dimensional runs will be presented in a future work (Sironi et al., \emph{in prep.}; see also Appendix \ref{sect:appendix_1D_vs_2D3D}, for the precursor energetics in 2D and 3D). We focus on the case of a cold pair-dominated plasma.
 
 There exists extensive literature on PIC modeling of the electromagnetic precursor wave in relativistic perpendicular shocks  \citep[e.g., ][]{langdon_88,gallant_92,hoshino_92,amato_arons_06, hoshino_08, sironi_spitkovsky_09, iwamoto_17, iwamoto_18}. Our work is motivated by the poor exploration of the extreme regime where the energy content of the plasma is dominated by magnetic fields, as it is supposedly the case in magnetar winds. In other words, we focus on magnetizations $\sigma\gtrsim 1$, where $\sigma$ is the ratio of upstream Poynting flux to kinetic energy flux. For $\sigma\gtrsim1$, 1D simulations are adequate, since we find that they agree well with multi-dimensional results (see Appendix \ref{sect:appendix_1D_vs_2D3D}, for the precursor energetics in 2D and 3D; also Sironi et al., \emph{in prep.}). We provide an extensive investigation of the dependence on the flow magnetization, from $\sigma=0.1$  to $\sigma=30$, with much longer simulations than previously reported, especially in the $\sigma\gtrsim1$ regime relevant for FRB sources. Previous works arguably never reached a steady state in simulations with $\sigma\gtrsim 1$. We employ several PIC codes (\textsc{Tristan-MP, Smilei} and \textsc{Shockapic}) to check for consistency, and thus confirm the robustness of our results.
  
 At $\sigma\gtrsim1$ the shock converts a fraction  $ f_\xi \approx 2 \times 10^{-3}/\sigma$ of the total incoming energy into the precursor wave, as measured in the post-shock (downstream) frame. In the shock rest frame, the efficiency is $f_\xi' \approx 7 \times 10^{-4}/\sigma^2$. The spectrum of the precursor wave is narrow-band, $\Delta \omega /\omega_{\rm peak} \lesssim 1-3$, with apparent but not dominant line-like features as many resonances concurrently contribute.
The peak frequency scales in the post-shock frame as $\omega_{\rm peak} \simeq 3\, \omega_{\rm p}\max[1,\sqrt{\sigma}]$, where $\omega_{\rm p}$ is the plasma frequency. In the pre-shock frame (which coincides with the observer frame, if the magnetar wind is non-relativistic),  this can be recast in a simpler form as $\omega^{\prime \prime}_{\rm peak} \approx 3 \gamma_{\rm s | u} \omega_{\rm p}$, where $\gamma_{\rm s|u}$ is the shock Lorentz factor in the upstream frame. At $\sigma\gtrsim1$, where our estimated $\omega_{\rm peak}$ differs from earlier works (that quoted $\omega_{\rm peak}\propto \sigma \omega_{\rm p}$, see  \citet{gallant_92},  rather than $\omega_{\rm peak}\propto \sqrt{\sigma} \omega_{\rm p}$ as we find) we see that the shock structure displays two solitons separated by a cavity, and the peak frequency of the spectrum corresponds to an eigenmode of the cavity. The efficiency and spectrum of the precursor wave do not depend on the bulk Lorentz factor of the pre-shock flow.

The paper is organized as follows. In section~\ref{sect:setup} we present the methods and the numerical setup. In section~\ref{sect:results} we discuss the main results, in the post-shock (downstream) rest frame. Section~\ref{sect:SRF} discusses the energy content of precursor waves in the frame of the shock front. In section~\ref{sect:discussion} we present the implications of our results for  FRB emission models, and we conclude in section~\ref{sect:conclusion}.

\section{Simulation methods and setup}
\label{sect:setup}

We use the particle-in-cell (PIC) codes \textsc{Tristan-MP} \citep{spitkovsky_05, sironi_spitkovsky_09}, \textsc{Smilei} \citep{SMILEI_paper}, and \textsc{Shockapic} \citep{plotnikov_18} to perform 1D3V simulations, where we retain one spatial direction, but all three components of velocities and electromagnetic fields. Mainly \textsc{Tristan-MP} and \textsc{Smilei} are employed for large simulations. The pseudo-spectral code \textsc{Shockapic} is used to check for consistency in shorter simulations. In the main body of the paper, we  present only results obtained with \textsc{Tristan-MP}, unless stated otherwise. In the Appendix~\ref{sect:appendix_codes}, we demonstrate the agreement between different codes across the whole range of $\sigma$ explored in this work.

 The use of a reduced 1D spatial geometry is justified in the limit of magnetically-dominated plasmas, as we will demonstrate  with 2D and 3D simulations in a forthcoming study (Sironi et al, \textit{in prep.}).
For lower magnetizations than explored here, i.e., $\sigma \lesssim 0.5$, \citet{iwamoto_17} found that the precursor wave energy is reduced by at most an order of magnitude in 2D simulations as compared to 1D. This difference is smaller or even negligible in the high magnetization limit explored here (see Appendix~\ref{sect:appendix_1D_vs_2D3D}).

 The shock is initialized using the common setup described in, e.g., \citet{spitkovsky_08b}, which we summarize here for completeness. The upstream flow, composed of electrons and positrons, drifts along the $-\hat{x}$ direction with a speed $-\beta_0c\hat{x}$. The corresponding bulk Lorentz factor is $\gamma_0=(1-\beta_0^2)^{-1/2}=10$, but we have also explored higher values of $\gamma_0$, up to $10^5$. The upstream pair plasma is cold, with thermal spread  $k_B T_0/(m_e c^2)=10^{-4}$.
 The pre-shock plasma carries a frozen-in magnetic field $B_0$ oriented along $z$ (so, $B_{z,0}=B_0$), i.e., perpendicular to the flow propagation, and  a motional electric field $E_{y,0}=-\beta_0 B_{z,0}$. The flow is reflected at a wall located at $x=0$. After some time (at least several cyclotron periods $\omega_{\rm c}^{-1}$), the shock front forms by magnetic reflection and steadily propagates along the $+\hat{x}$ direction with a speed that is in good agreement with the Rankine-Hugoniot conditions. The resulting simulation frame coincides with the frame where the downstream plasma is at rest (downstream rest frame; DRF).
 
The shock physics is sensitive to the upstream magnetization $\sigma$, which we define as the ratio of Poynting  to kinetic energy flux
\begin{equation}
\sigma = {B_{z,0}^2 \over 8 \pi \gamma_0 N_0 m_e c^2} = \left( {\omega_{\rm c} \over  \omega_{\rm p}} \right)^2 \, ,
\label{eq:define_sigma}
\end{equation}
and we vary from $\sigma=0.1$ up to $\sigma=30$.
Here, $N_0$ is the number density of upstream electrons (the overall particle number density is then $2 N_0$), $m_e$ is the electron (or positron) mass, and $c$ is the speed of light in vacuum. We also define the typical gyro-frequency as $\omega_{\rm c}= |q| B_0 / (\gamma_0 m_e c)$ and the plasma frequency as $\omega_{\rm p}=[8\pi N_0 q^2 / (\gamma_0 m_e)]^{1/2}$, where $q$ is the elementary electric charge. Both quantities are based on the upstream values of magnetic field and plasma density measured in the simulation frame.

Typical numerical parameters used with \textsc{Tristan-MP} are:
\begin{itemize}
\item The skin depth is well resolved with $c/\omega_{\rm p}=100 \,\Delta$, where $\Delta$ is the grid size. This ensures that the typical particle gyro-radius $\simeq \sigma^{-1/2}\,c/\omega_{\rm p}$ is well resolved even for the largest magnetization $\sigma=30$ explored in this study. A good spatial resolution is also essential to capture the high-frequency part of the spectrum of precursor waves \citep[see also][ for a discussion on the required spatial resolution]{iwamoto_17}.
\item  The simulation time-step is defined as $c\Delta t=0.5\, \Delta$, corresponding to a time resolution of $5\times 10^{-3}\omega_{\rm p}^{-1}$.
\item  The number of particles per cell  initialized in the upstream plasma is $N_{ppc}=64$ per species (values between 20 and 200 were tested with no appreciable differences).
\item The simulation is evolved up to $T_{\rm sim} \simeq 1.5\times10^3 \omega_{\rm p}^{-1}$ for $\sigma<1$ and for longer times (up to $2 \times 10^4 \omega_{\rm p}^{-1}$)  for $\sigma \gg 1$.  This is required in order to reach a steady state in which the precursor emission maintains a constant amplitude (see section~\ref{subsect:wave_energy_sigmas}).
\end{itemize}
In the Appendix~\ref{sect:appendix_codes} we also report the typical simulation parameters for the other two PIC codes used in this study. They are not presented here since  in the following sections we mainly discuss the results obtained with \textsc{Tristan-MP}.


\section{Results}
\label{sect:results}

In this section we explore the physics of relativistic highly magnetized electron-positron shocks,  focusing on the properties of the electromagnetic precursor. In section~\ref{subsect:shock_struct} we discuss the typical shock structure and show the presence of electromagnetic precursor waves. In sections~\ref{subsect:wave_energy_sigmas} and \ref{subsect:spectrum} we show the dependence on $\sigma$ of the precursor wave intensity and spectrum, respectively. The wave strength parameter is discussed in section~\ref{subsect:strength_parameter}. 

\subsection{Shock layer structure}
\label{subsect:shock_struct}

\begin{figure*}
	\begin{center}
		\includegraphics[width=0.95\textwidth,angle=0]{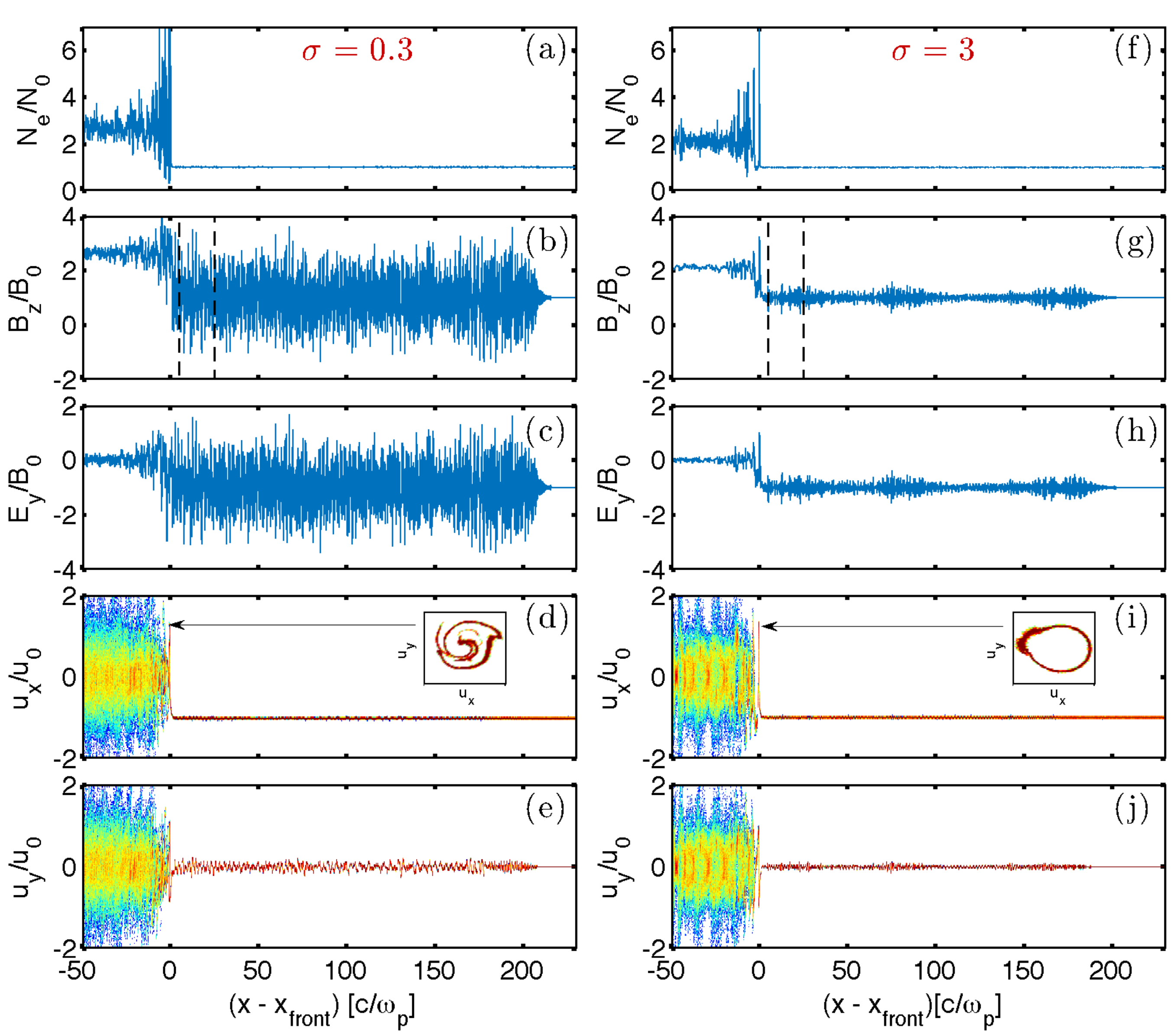}
		\caption{Structure of the shock layer  from 1D PIC simulations of $\sigma=0.3$ (left colum) and $\sigma=3$ (right column) shocks. From top to bottom it is shown: the electron number density $N_e/N_0$ (panels a and f), the transverse magnetic field $B_z/B_0$ (panels b and g), the motional transverse electric field $E_y/B_0$ (panels c and h), the longitudinal phase space of positrons $x-u_x$ (panels d and i), and the transverse phase space of positrons $x-u_y$ (panels e and j), respectively. Here, $u_x=\gamma \beta_x$ and $u_y=\gamma \beta_x$ are the components of the dimensionless four-velocity. Small insets in the upper right side of  panels (d) and (i) show the particle distribution in momentum space $u_x-u_y$ at the location of the shock front, as indicated by the arrows. The shock front is located at $x-x_{\rm front}=0$; it propagates in the $+\hat{x}$ direction. The upstream flow is on the positive $x-x_{\rm front}>0$ side and the downstream plasma is on the negative $x-x_{\rm front}<0$ side.}
		\label{fig:1D_structure}
	\end{center}
\end{figure*}

As shown in several works using 1D simulations \citep{langdon_88, gallant_92, amato_arons_06, lyubarsky_06, hoshino_08}, 2D simulations \citep{sironi_spitkovsky_09, sironi_spitkovsky_11, iwamoto_17, iwamoto_18, plotnikov_18} and 3D simulations \citep{spitkovsky_05, sironi_13}, highly magnetized perpendicular relativistic shocks form by magnetic reflection and generate a strong electromagnetic wave propagating from the shock  into the upstream region. For high magnetizations (typically, $\sigma\gtrsim0.01$), the transverse Weibel filamentation instability, which dominates for $\sigma \lesssim 10^{-3}$, plays no significant role in shaping the shock structure. This partly justifies the 1D approach adopted here. 

In Figure~\ref{fig:1D_structure} we present the structure of the shock transition region for two representative magnetizations. The left column (panels a-e) presents a shock with $\sigma=0.3$ and the right column (panels f-j) corresponds to $\sigma=3$. The timespan of the simulations is long enough to reach a stationary state: we show results at $\omega_{\rm p}t=540$ for $\sigma=0.3$ and at $\omega_{\rm p}t=1800$ for $\sigma=3$. From top to bottom we show the electron number density $N_e/N_0$ (panels a and f), the transverse magnetic field $B_z/B_0$ (b and g), the transverse electric field $E_y/B_0$ (c and h), the longitudinal positron phase space $x-u_x$ (d and i), and the transverse positron phase space $x-u_y$ (e and j). Here, we define $u_\alpha=\gamma \beta_\alpha$ as the dimensionless four-velocity. The phase space of electrons is identical to the one of positrons in virtue of mass symmetry, except for the opposite sign in variations of $u_y$. The electrostatic field $E_x$ is not plotted since it is completely negligible in pair plasmas (we have systematically checked this conclusion). The vertical dashed lines in panels (b) and (g) delimit the region where we have extracted the wave properties, such as amplitude and spectrum,  that will be discussed in the sections below. Small insets in the upper right side of panels (d) and (i) show the particle distribution in momentum space $u_x-u_y$ at the location of the shock front.

The shock front is located at $x-x_{\rm front}=0$ in Figure~\ref{fig:1D_structure}. The upstream flow is on the positive side $(x-x_{\rm front}>0)$ and the downstream plasma is on the negative side $(x-x_{\rm front}<0)$. The existence of a well developed shock is confirmed by the jump in the electron number density and in the $B_z$ field at the front location. The shock front itself exhibits a soliton-like structure \citep[see, e.g.,][]{alsop_arons_88}, where the particle distribution forms a semi-coherent cold ring in momentum space (see insets in panels d and i). The presence of a large amplitude electromagnetic precursor wave is evidenced in the upstream region of the $B_z/B_0$ and $E_y/B_0$ panels, for both magnetizations (see the ripples in the  $x-x_{\rm front}>0$ region). The precursor wave amplitude is larger for $\sigma=0.3$ than for $\sigma=3$. This wave is steadily emitted from the shock front and is \emph{linearly polarized}. The wave vector $\bmath{k}$ lies along the shock direction of propagation (i.e., along $x$), the fluctuating magnetic field is along  $z$  (i.e., along the same direction as the upstream field $\bmath{B}_0=B_0\hat{z}$), and the fluctuating electric field is perpendicular to both $\bmath{k}$ and $\bmath{B}_0$. The wave is then identified with the extraordinary mode (X-mode). The phase velocity of the wave is slightly superluminal, as expected for X-mode propagation in a plasma, while its electromagnetic nature is confirmed by the fact that the space-averaged $\langle \delta B_z^2 \rangle = \langle \delta B_z \delta E_y \rangle$.

We note that the field-aligned component of the particle momentum $u_z$ is not affected by the shock. The incoming particles are efficiently isotropized in the $xy$ plane perpendicular to the field, but the post-shock particle distribution remains largely confined to this plane. It follows that the downstream effective adiabatic index corresponds to a 2D relativistically hot gas, $\Gamma_{\rm ad}=3/2$, instead of $4/3$ if the downstream plasma were isotropic in all momentum directions. The lack of isotropization is due to the fact that in a $\sigma \gg 1$ flow (with downstream plasma magnetically dominated) it will be harder for the plasma to exceed the threshold for velocity-space instabilities that feed off the particle temperature anisotropy. For example, the plasma will go unstable via the mirror mode if the temperature anisotropy is above a threshold that scales as $\propto \sigma$, which is harder to exceed at higher magnetizations. In addition, the 1D spatial geometry employed here will further  suppress the growth of field-aligned modes leading to momentum isotropization.

The downstream particle energy spectrum (not shown) resembles a 2D Maxwell-J\"uttner distribution whose temperature is slightly lower than the one expected from the Rankine-Hugoniot jump conditions (the difference is due to the energy transferred to the precusor waves). No non-thermal tail is observed for the runs presented in this work.
This is in agreement with the inefficiency of particle acceleration expected at relativistic strongly magnetized perpendicular shocks \citep[][]{sironi_spitkovsky_09, lemoine_pelletier_10, sironi_13, sironi_15, pelletier_17, iwamoto_17,plotnikov_18}.

\subsection{Precursor wave  energy}
\label{subsect:wave_energy_sigmas}

We now focus on the dependence of the precursor properties, and specifically of its amplitude, on the upstream magnetization $\sigma$.
The dependence on the upstream bulk Lorentz factor $\gamma_0$ will be discussed in the last part of this subsection.

\subsubsection{Temporal evolution of the precursor wave}
After an initial transient --- whose duration depends on the upstream  magnetization, as we show below --- the intensity of the precursor wave settles to its asymptotic value. We measure the wave intensity in a region between 5 and 25 $c/\omega_{\rm p}$ ahead of the shock front:\footnote{This region is delimited by vertical black dashed lines in panels (b) and (g) of Figure~~\ref{fig:1D_structure}.} $5\,c/\omega_{\rm p} < x-x_{\rm front}<25\, c/\omega_{\rm p}$. This region is far enough from the shock not to be affected by the front structure itself, and it contains a large number of precursor wavelengths so that we can obtain a solid measure of the precursor average properties. 

\begin{figure}
\begin{center}
\includegraphics[width=0.45\textwidth,angle=0]{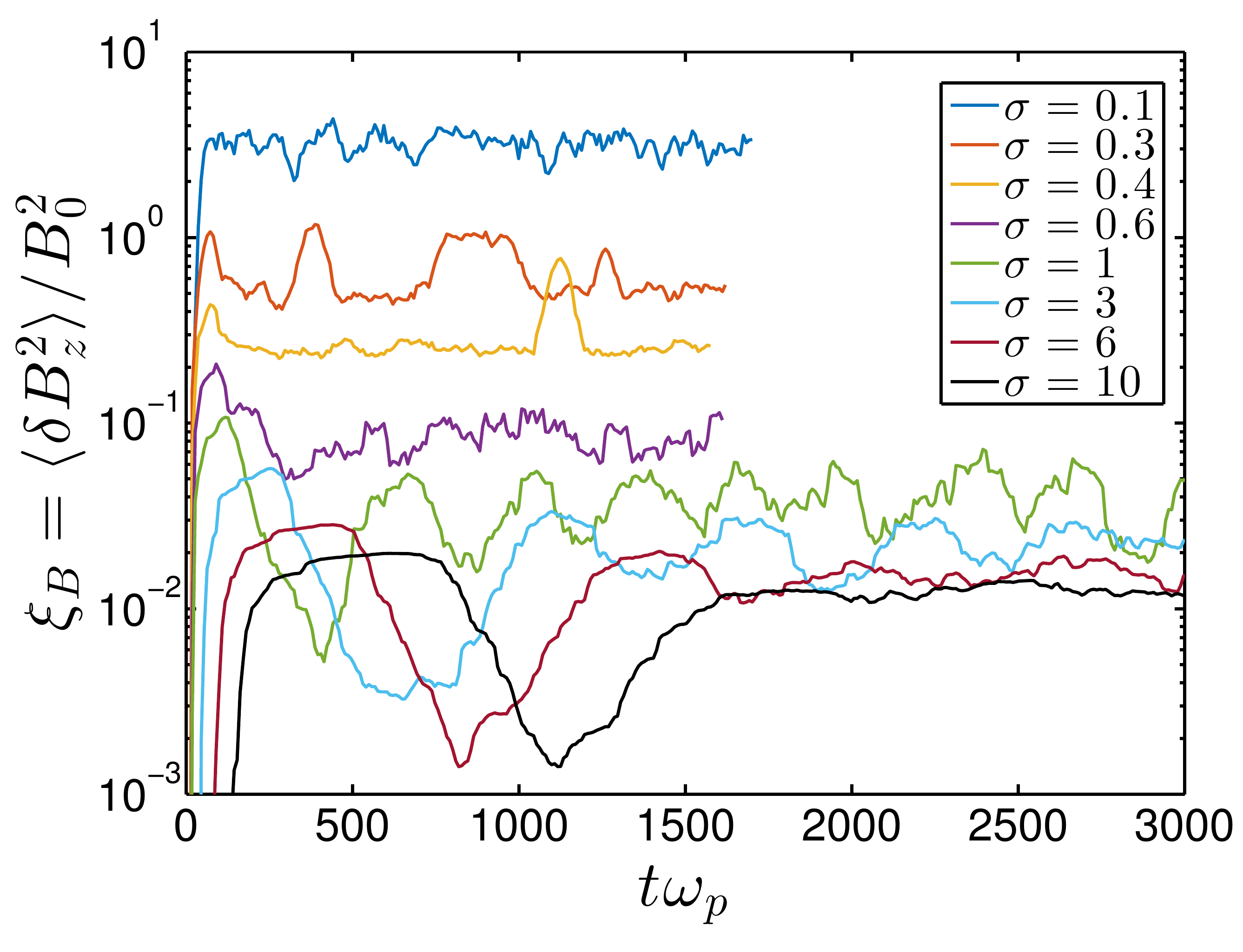}
\caption{Time evolution of the normalized precursor wave energy, $\xi_B= {\langle \delta B_z^2 \rangle / B_0^2} $, for different values of the upstream magnetization $\sigma$. Lines of different color correspond to a given $\sigma$ going from $0.1$ (blue line) to $10$ (black line). The precursor wave energy was extracted from a $20\, c/\omega_{\rm p}$-wide slab located at $5\,c/\omega_{\rm p} < x-x_{\rm front}<25\, c/\omega_{\rm p}$.}
\label{fig:xiB_time_evolution}
\end{center}
\end{figure}

The wave intensity is then calculated as the spatial average
\begin{equation}
\langle \delta B_z^2 \rangle =\langle (B_z-B_0)^2 \rangle \, .
\end{equation}
In Figure~\ref{fig:xiB_time_evolution} we show  for different magnetizations the time evolution of the normalized wave intensity, defined as
\begin{equation}
\xi_B = {\langle \delta B_z^2 \rangle \over B_0^2} = {\langle \delta B_z \delta E_y \rangle \over B_0^2}~,
\label{eq:xiB_definition}
\end{equation}
where $\delta E_y=E_y-E_{y,0}=E_y+\beta_0B_0$.
Different solid lines correspond to different values of the magnetization, from $\sigma=0.1$ (blue line) to $\sigma=10$ (black line), as indicated in the legend. By computing the temporal variation of the wave intensity, we can assess when the precursor wave has reached a steady state.
Figure~\ref{fig:xiB_time_evolution} shows that:

\begin{itemize}
\item With increasing $\sigma$, a longer time is required for the precursor to settle at its time-asymptotic state. This is due to the combination of two effects. First,  the shock velocity increases from $\beta_{\rm s|d}=0.476$ for $\sigma=0.1$ to $\beta_{\rm s|d}=0.963$ for $\sigma=10$, so at higher magnetizations it takes more time for the precursor wave, propagating at $\beta_{\rm wave} \simeq 1$, to detach from the shock front. Second, there is some interaction occurring between the wave and the upstream plasma, which initially causes a drop in wave efficiency (e.g., at $\ompt\sim 1100$ in the black line of Figure~\ref{fig:xiB_time_evolution}).
The time required for the wave to self-regulate and settle to a steady state, following this drop, is longer for higher $\sigma$ (e.g., compare green and black lines in Figure~\ref{fig:xiB_time_evolution}).
\item The steady-state value of the normalized wave energy $\xi_B$ decreases with increasing magnetization and for $\sigma \gg 1$ (cyan, brown and black lines) it approaches a constant value $\xi_B \simeq 0.01$.
\item For $\sigma=0.3$ (orange) and $\sigma=0.4$ (yellow), which will be called ``transition cases'' in the following, the wave intensity varies between periods of high efficiency and phases of low efficiency.
\item All the simulations have been evolved for long enough to reach a quasi-stationary state. For the largest explored magnetization $\sigma=30$, the simulation was advanced beyond $2\times 10^4~\omega_{\rm p}^{-1}$ (this case is not shown in the figure but reported in subsequent figures).
\end{itemize}

Once the wave intensity has settled to a steady state, we have extracted a number of wave properties, such as the energy, the spectrum (peak frequency, low-frequency cutoff, spectral width), and the wave strength parameter, as we now describe.

\subsubsection{Dependence on the upstream magnetization}

In the previous section we have defined the energy fraction in upstream field fluctuations $\xi_B$  (see Eq.~\ref{eq:xiB_definition}), as the ratio of the precursor wave magnetic energy to the background field energy. In order to get a global idea of the energetics, we also need to complement it with a parameter that quantifies the energy fraction of the incoming plasma (including both kinetic and electromagnetic content) that is radiated from the shock front in the form of precursor waves. When the electromagnetic field fluctuations induced by the precursor are taken into account in the jump conditions 
across the shock, the energy conservation equation expressed in the simulation frame is \citep{gallant_92, plotnikov_18}:
\begin{eqnarray}
\gamma_0^2  \left( \beta_0+\beta_{\rm s|d} \right) \left( w_u+{b_{0,u}^2 \over 4\pi} \right) - \left( 1 - \beta_{\rm s|d} \right) {\delta B_{u}^2 \over 4\pi}  = \nonumber \\ 
\beta_{\rm s|d} \left( w_d - p_d + {b_{0,d}^2 \over 8 \pi} +{\delta B_d^2\over 4 \pi} \right) \, ,
\label{eq:energy_cons_with_waves_DRF}
\end{eqnarray}

where the subscripts `$u$' and `$d$' refer to the upstream and downstream regions, respectively. Here, $w_i$ and $b_{0,i}$ are respectively the fluid enthalpy density and mean magnetic field, both measured in the fluid rest-frame. As above, $\beta_{\rm s | d}$ is the velocity of the shock front as measured in the downstream frame of the simulations. The fluctuating components $\delta B_{u}$ and $\delta B_{d}$ are measured  in the simulation frame. We have made the approximation of negligible thermal pressure upstream (strong shock limit) and we have assumed that electrostatic effects are negligible both upstream and downstream ($\delta E_{x} \approx 0$). The latter approximation is fully supported by the simulations and, more fundamentally, by the fact that space-charge effects  are expected to be negligible in pair plasmas. 
The strong shock limit means that the upstream plasma pressure can be neglected and the upstream fluid enthalpy density (in the fluid rest-frame) is then $w_u=n_u m_e c^2$, where $n_u$ is the upstream plasma proper density. The mean upstream magnetic field $b_{0,u}$ in the upstream frame is related to the pre-shock magnetic field $B_{z,0}$ in the simulation frame via a Lorentz boost: $ B_{z,0} =  \gamma_0 b_{0,u}$. Hence, the magnetization parameter can be rewritten as $\sigma= b_{0,u}^2 / (4\pi w_u)$ and we note that $\delta B_u^2 /b_{0,u}^2=\gamma_0^2 \delta B_u^2 /B_{z,0}^2 =\gamma_0^2 \xi_B$, because the fluctuating part is measured directly in the simulation frame. 

As we focus on the precursor wave propagating upstream, here we only consider the left hand side of equation~\ref{eq:energy_cons_with_waves_DRF}.\footnote{Since we forego the discussion of the downstream part of the energy conservation equation, an interested reader will find details in the aforementioned works \citep{gallant_92, plotnikov_18}.}
We find that the fraction of total incoming energy (including both particle and electromagnetic  contributions) that is channeled into the precursor wave can be expressed as
 \begin{equation}
f_\xi = \xi_B \left({\sigma \over 1+\sigma} \right) \left( {1-\beta_{\rm s|d} \over \beta_0 + \beta_{\rm s|d}} \right) \, ,
\label{eq:fxi_DRF}
\end{equation}
as seen from the DRF. In the following, $f_\xi$ will be identified as the ``energy fraction parameter''. It is also convenient to define the fraction of incoming particle kinetic energy that is converted into precursor emission
 \begin{equation}
g_\xi=f_\xi\, (1+\sigma)~.
\end{equation}

\begin{figure}
\begin{center}
\includegraphics[width=0.45\textwidth,angle=0]{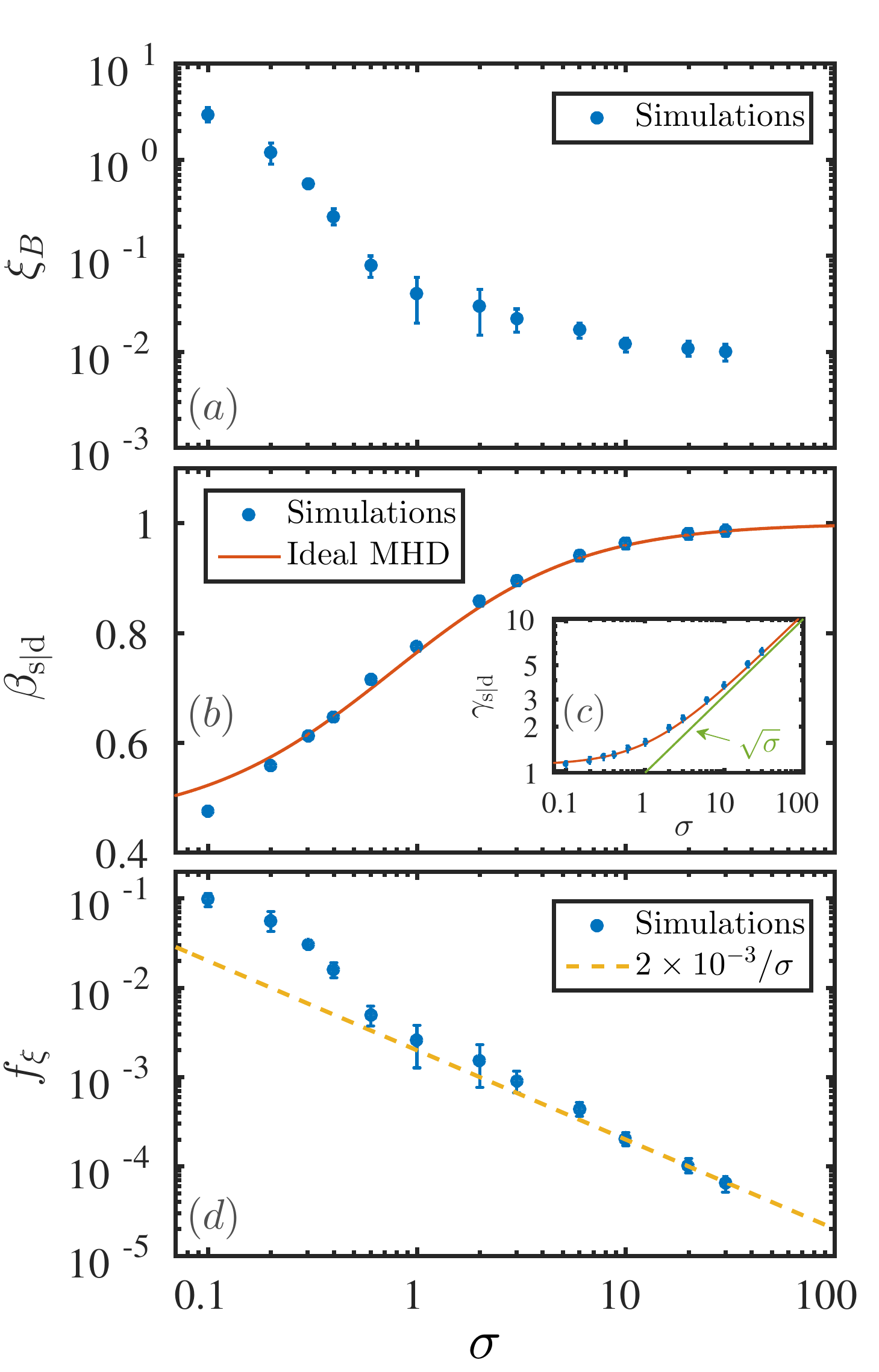}
\caption{Dependence on $\sigma$ of the time-asymptotic values of the normalized wave intensity $\xi_B$ (panel a), of the shock front speed $\beta_{\rm s | d}$ (panel b),  of the corresponding Lorentz factor $\gamma_{\rm s | d}$ (panel c), and of the energy fraction parameter $f_\xi$ (panel d). The red lines in panels (b) and (c) show the theoretical expectation based on ideal MHD jump conditions (which do not include effects from the precursor waves). The green line in panel (c) shows the asymptotic scaling $\gamma_{\rm s | d} =\sqrt{\sigma}$, expected for $\sigma \gg 1$. The yellow dashed line in panel (d) follows the scaling $f_\xi = 2\times 10^{-3}\sigma^{-1}$. }
\label{fig:xiB_sigma_3panel}
\end{center}
\end{figure}

The time-asymptotic values of $\xi_B$, $\beta_{\rm s | d}$, and $f_\xi$ measured in our simulations are presented in panels (a), (b) and (d) of Figure~\ref{fig:xiB_sigma_3panel}, as a function of magnetization. Error bars indicate the standard deviation of our time measurements.  
As regard to $\xi_B$, we observe a rapid decrease from  $\xi_B\simeq 3.53$ at $\sigma=0.1$ down to   $\xi_B\simeq 0.04$ at $\sigma=1$. The inflection point of the transition occurs at $\sigma \simeq 0.35$. This decrease  accompanies a change in the shock front structure that for $\sigma>1$ presents a coherent soliton-like shape (compare left and right columns in figure~\ref{fig:1D_structure} at the shock). For $\sigma>1$, $\xi_B$ slowly decreases and eventually approaches a constant value $\xi_B\simeq10^{-2}$.  

Concerning the shock front speed and its corresponding bulk Lorentz factor, $\beta_{\rm s|d}$ and $\gamma_{\rm s|d}$, panels (b) and (c) demonstrate an excellent agreement between our measured values, plotted with blue symbols, and the predictions of ideal MHD jump conditions \citep[e.g., Appendix B of][]{plotnikov_18}, as indicated by the red solid line. The front speed increases from $\beta_{\rm s|d}=0.476c$ for $\sigma=0.1$ up to $\beta_{\rm s|d}=0.987c$ for $\sigma=30$. The Lorentz factor of the shock front tends asymptotically to  $\gamma_{\rm s|d} = \sqrt{\sigma}$, for $\sigma \gg 1$. We remark that the MHD equations used here to derive the jump conditions do not incorporate modifications due to the precursor wave. The accurate agreement of our results with ideal MHD jump conditions for $\sigma \gg 0.1$ is then due to the fact that at high magnetizations the precursor wave is relatively weak, and it does not have an appreciable dynamical effect on the shock. In contrast, in the case when the precursor wave is the strongest, $\sigma=0.1$, the agreement is the worst, because the emission of the large amplitude wave can slow down the shock front, as compared to the ideal MHD prediction.

The dependence on $\sigma$ of the energy fraction parameter $f_\xi$ is presented in panel (d) of figure~\ref{fig:xiB_sigma_3panel}. It was calculated by plugging the values from panels (a) and (b) into equation~\ref{eq:fxi_DRF}. It shows that the energy fraction in the precursor wave decreases from $10$\% for $\sigma=0.1$ down to $0.0065$\% for $\sigma=30$. The dashed orange line follows the empirical scaling $f_\xi =2\times 10^{-3}/\sigma$ that satisfactorily fits our measured values in the $\sigma>1$ range.
 The most noticeable result of this panel is that we observe a well-defined scaling $f_{\xi} \propto \sigma^{-1}$. This result arises from the fact that  for $\sigma \gg 1$, the normalized wave intensity $\xi_B$ is roughly constant and $\beta_{\rm s|d} \simeq 1-1/(2\sigma)$. It follows that in the limit $\sigma\gg1$ the precursor wave carries a constant fraction of the incoming particle kinetic energy, i.e., $g_\xi\simeq 2\times 10^{-3}$.
 
Let us emphasize, however, that this $\sigma$-dependence of $f_\xi$ and $g_\xi$ is derived in the DRF (simulation frame). This dependence will be different in the shock front rest frame, since the front moves with ultra-relativistic speeds for $\sigma \gg 1$. This point will be further discussed  in section~\ref{sect:SRF}. 

\subsubsection{Dependence on the upstream bulk Lorentz factor}
\label{subsubsect:depend_on_gamma0}

\begin{figure}
\begin{center}
\includegraphics[width=0.49\textwidth,angle=0]{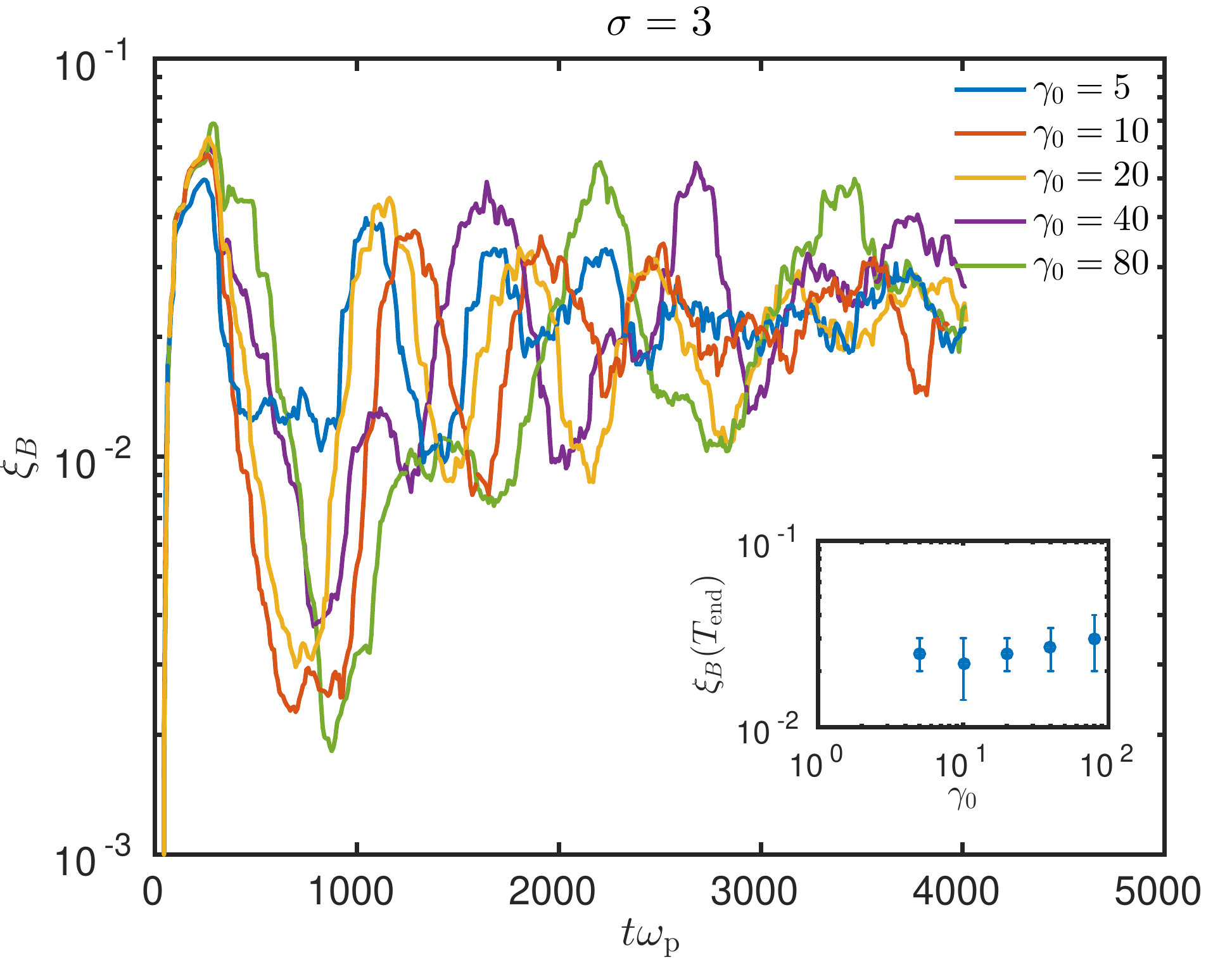}
\caption{Time evolution of the normalized wave intensity $\xi_B$ for different values of the upstream flow Lorentz factor $\gamma_0$, at fixed $\sigma=3$. Blue, red, orange, violet, and green lines correspond to $\gamma_0=5,\,10,\,20,\,40$, and $80$, respectively. The time-asymptotic values of $\xi_B$, with corresponding error bars, are plotted in the inset at the lower right corner of the figure. Within the error bars, $\xi_B$ is nearly independent from $\gamma_0$.}
\label{fig:gamma0_dependence}
\end{center}
\end{figure}

So far, we have investigated the dependence of the precursor intensity on $\sigma$, for a fixed choice of the upstream flow Lorentz factor $\gamma_0=10$. Here, we demonstrate that $\xi_B$ is essentially independent from $\gamma_0$, for any value of $\sigma$.  Let us first consider the dependence at a fixed $\sigma$. In Figure~\ref{fig:gamma0_dependence} we show the time evolution of the precursor wave energy for $\sigma=3$, when varying $\gamma_0$ from $5$ to $80$. Lines of different color correspond to different values of $\gamma_0$. Despite large oscillations in time, it appears that $\xi_B$ converges to the same value, regardless of $\gamma_0$. The time-asymptotic values of $\xi_B$, with corresponding error bars, are plotted in the figure inset. Within the error bars, we can assert that there is no obvious dependence on $\gamma_0$. 

In order to generalize this conclusion to any $\sigma$, it is worth noting that in the seminal study of \citet{gallant_92}, two very different values of the bulk Lorentz factor ($\gamma_0=40$ and $10^6$) were used, for a range of $\sigma \in [10^{-3},5]$. The authors did not notice any dependence on $\gamma_0$. Also, \citet{iwamoto_17} performed 1D simulations with $\gamma_0=40$ and explored $\sigma$ values between $10^{-3}$ and $0.5$, finding similar values as in \citet{gallant_92}. 

In the Appendix~\ref{sect:appendix_codes}, Figure~\ref{fig:codes_compar_xiB_litterature} shows the values of $\xi_B$ obtained for $\sigma \in [10^{-3},1]$ (horizontal axis) and for $\gamma_0$ ranging from $10$ to $10^6$ (different datasets). This figure shows that in the low magnetization regime $\sigma \in [10^{-3},0.3]$, the normalized wave intensity $\xi_B$ is nearly independent from $\gamma_0$.
In the range $\sigma \in [0.3,1]$ there is a larger scatter among different datasets (which employ different $\gamma_0$). This range of magnetizations corresponds to the transition cases (see Figure~\ref{fig:xiB_time_evolution}). The most plausible reason for the discrepancy among different datasets is that the simulations from earlier studies were not evolved long enough in order to reach the asymptotic state of the transition cases, so the value of $\xi_B$ was not yet stabilized (see figure~\ref{fig:xiB_time_evolution}). In fact, Figure~\ref{fig:gamma0_dependence} shows that even at $\sigma>1$ the time-asymptotic value of $\xi_B$ is insensitive to the flow Lorentz factor.


\subsection{Precursor spectrum}
\label{subsect:spectrum}

After discussing the wave energy, we now address the dependence on $\sigma$ of the precursor spectrum and of the typical wavelength of the emission. Our results will be presented in the downstream frame of the simulations.
It is important to note, however, that the wave propagates in the upstream plasma and that its emitter is the shock front. Both move with respect to the simulation frame. 
Hence, when comparing simulation results with the expected scalings, we need to consider the wave dispersion relation first in the upstream frame, and then transform it to the DRF. Also, the typical emission frequency is most naturally estimated in the shock rest frame, and then it should be transformed to the DRF in order to compare with simulation results. 

In this section, we first present basic analytical considerations and then we compare them with our simulation results. A special feature of $\sigma >1$ shocks, where a density and magnetic field cavity is observed in the front structure, is discussed at the end of this section. As we argue, the cavity is instrumental in setting the precursor power and determining its dominant frequency.

\subsubsection{Basic considerations}
As discussed above, the precursor wave possesses X-mode (extraordinary-mode) polarization, such that its wave vector is perpendicular to $\bmath{B}_0$, its fluctuating magnetic field is parallel to $\bmath{B}_0$, and its fluctuating electric field is perpendicular to both $\bmath{k}$ and $\bmath{B}_0$. Some basic properties of the extraordinary mode in the context of the shock emission were derived by \citet{gallant_92} and \citet{iwamoto_17}. We reproduce here their estimations for completeness.

The dispersion relation of the extraordinary mode in the frame where the background plasma is at rest reads \citep[see, e.g.,][]{hoshino_91}
\begin{eqnarray}
{k^{\prime \prime 2} c^2 \over \omega^{\prime \prime 2}} &=& 1 - {\omega_{\rm p} ^{\prime \prime 2} \over  \omega^{\prime \prime 2} - \omega_{\rm c} ^{\prime \prime 2}} = 1 - {\omega_{\rm p} ^{\prime \prime 2} \over \omega^{\prime \prime 2} -\sigma \omega_{\rm p} ^{\prime \prime 2}} \, ,
\end{eqnarray}
where double primed quantities are measured in the upstream rest frame (URF). Using Lorentz transformations for $\omega$ and $k$ and in the limit $\gamma_0^2 \gg \sigma$, the dispersion relation in the DRF becomes
\begin{equation}
k^2 c^2 \simeq \omega^2-\omega_{\rm p}^2 \, .
\label{eq:dispersion_DRF}
\end{equation}
Interestingly, as long as $\gamma_0^2 \gg \sigma$, this is identical to the dispersion relation of a simple electromagnetic wave propagating in an unmagnetized plasma.

The motion of the shock front imposes a cutoff frequency below which the wave cannot escape into the upstream medium. It follows that little or no power should be observed in the upstream precursor spectrum below the cutoff frequency. This cutoff frequency is obtained by equating the group velocity of the wave, ${\rm d} \omega / {\rm d} k$, with the shock front velocity as:
\begin{equation}
c \sqrt{1-{\omega_{\rm p}^2 \over \omega^2}} = \beta_{\rm s | d} c \, .
\end{equation}
This relation leads to the cutoff frequency and wavelength
\begin{eqnarray}
\omega_{\rm cutoff}& =& \gamma_{\rm s | d} \omega_{\rm p} \label{eq:omega_cutoff} \\
\lambda_{\rm cutoff}& =& {2\pi c \over \gamma_{\rm s | d} \beta_{\rm s | d} \omega_{\rm p}} \, .
 \label{eq:lambda_cutoff}
\end{eqnarray}

As regard to the characteristic frequency of the precursor wave, the most natural assumption is that it corresponds to the collective cyclotron motion of the bunching particles at the shock front, which we now evaluate.
First, the magnetic field at the shock can be roughly estimated by assuming that, in the shock frame, all the momentum of the incoming particles is stored in the magnetic field at that point \citep{alsop_arons_88}:
\begin{equation}
{B_{\rm sh}^{\prime} \over B_0^{\prime}} \approx \sqrt{1+{2\over \sigma}} \, ,
\label{eq:Bmax_front}
\end{equation}
where primed quantities are measured in the shock rest frame (SRF). More detailed considerations on the soliton structure of the shock as presented by \citet{alsop_arons_88}, give a similar expression for $B_{\rm sh}^{\prime}$.
For particles with Lorentz factors comparable to the upstream bulk Lorentz factor, the
ratio of the expected emission frequency (which we label ``sol'' since it is emitted by the soliton at the shock) to the upstream cyclotron frequency is then equal to the magnetic field enhancement ratio, $\omega_{\rm c, sol}^{\prime}/\omega_{\rm c} =  B_{\rm sh}^{\prime} / B_0^{\prime}$.\footnote{There is no prime on the upstream cyclotron frequency as it is Lorentz-invariant for perpendicular shocks.}
Lorentz transforming to the DRF ($\omega_{\rm c, sol}^{\prime} \to \omega_{\rm c, sol}$) and using
the dispersion relation in  Eq.~\ref{eq:dispersion_DRF} leads to
\begin{equation}
\omega_{\rm c, sol} \approx \left(\sqrt{\sigma+2} + \sqrt{\sigma + 2-\beta_{\rm s |d}^2} \right) \gamma_{\rm s | d} \omega_{\rm p}  \, .
\label{eq:omega_peak}
\end{equation}

Based on these arguments, we expect the precursor spectrum to exhibit a low-frequency cutoff at $\omega_{\rm cutoff}$ and prominent line-like features at $\omega_{\rm c, sol}$ and its harmonics. As we show below, where we compare these scalings with our simulation results, for $\sigma>1$ the predicted $\omega_{\rm c, sol}$ systematically over-estimates the observed peak frequency $\omega_{\rm peak}$. In section~\ref{subsubsect:cavity}, we  propose a new model for the  precursor peak frequency  in the high-magnetization regime, and we show that it is in good agreement with our simulation results.

\subsubsection{Spectrum dependence on the upstream magnetization}

To characterize the spectrum of the precursor wave, we have employed two complementary diagnostics, one spatial and one temporal. They were used to construct the wavenumber spectrum ($k$-spectrum) and the frequency spectrum ($\omega$-spectrum), respectively. 

The wavenumber spectrum was calculated by extracting the spatial profile of $B_z(x)-B_{z,0}$ in the region located at $5\, c/\omega_{\rm p} < x-x_{\rm front} < 105\, c/\omega_{\rm p}$, at a time when the precursor has reached the steady state, and then computing its Fourier transform.  The frequency spectrum was constructed by recording the temporal variation of $B_z(t)-B_{z,0}$ at one selected grid point in the upstream region, during a time interval of $100\, \omega_{\rm p}^{-1}$, and then calculating its Fourier transform. The spatial window for the $k$-spectrum and the time interval for the $\omega$-spectrum are chosen so that  
roughly the same segment of the precursor wave was analyzed in the two cases.

\begin{figure}
\begin{center}
\includegraphics[width=0.47\textwidth,angle=0]{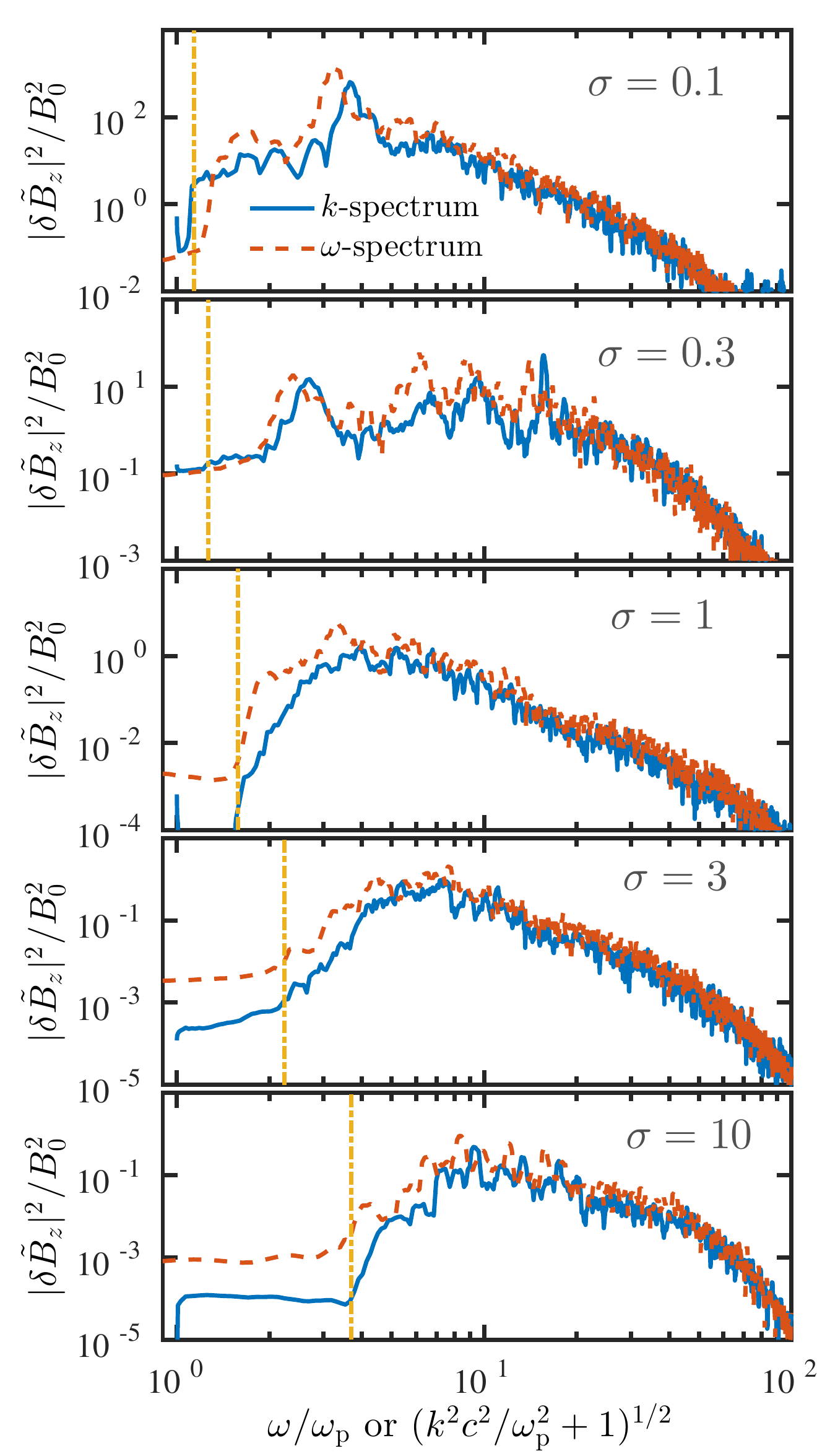}
\caption{Spectrum of the precursor wave for different $\sigma$. Five representative cases are shown from top to bottom, $\sigma=0.1, 0.3, 1, 3,$ and $10$, respectively.  Blue solid lines in each panel show the spectrum in $k$-space and red dashed lines show the same in $\omega$-space. The horizontal axis shows $\omega/\omega_{\rm p}$ for the $\omega$-spectrum. For the $k$-spectrum, the choice for the horizontal axis is motivated by the dispersion relation in Eq.~\ref{eq:dispersion_DRF}, so that the $k$-spectrum should nearly overlap with the corresponding $\omega$-spectrum. The method to compute the spectra is described in the main text. All the spectra are normalized as $\int |\delta \tilde{B}_z(k)|^2 /B_0^2\, {\rm d}k  =\int |\delta \tilde{B}_z(\omega)|^2 /B_0^2\, {\rm d}\omega = \xi_B$. The orange vertical lines mark the position of the expected low-frequency cutoff, as given by Eq.~\ref{eq:omega_cutoff}.}
\label{fig:spectrum_five_sigmas}
\end{center}
\end{figure}

In Figure~\ref{fig:spectrum_five_sigmas} we present the spectrum of the precursor wave for different $\sigma$. Five representative cases are shown from top to bottom, $\sigma=0.1, 0.3, 1, 3,$ and $10$, respectively. Each panel contains the $k$-spectrum, plotted using blue solid lines, and the $\omega$-spectrum, plotted using red dashed lines. For the $\omega$-spectrum, the horizontal axis shows $\omega/\omega_{\rm p}$, whereas for the $k$-spectrum we take  $(k^2c^2/\omega_{\rm p}^2+1)^{1/2}$. Due to this choice, and given the dispersion relation in Eq.~\ref{eq:dispersion_DRF}, each wavenumber spectrum should nearly overlap with the corresponding frequency spectrum, as it is indeed the case. The spectra were normalized such that $\int |\delta \tilde{B}_z(k)|^2 /B_0^2\, {\rm d}k  =\int |\delta \tilde{B}_z(\omega)|^2 /B_0^2\, {\rm d}\omega = \xi_B$.

In each spectrum, the power drops rapidly below the  cutoff frequency given by Eq.~\ref{eq:omega_cutoff}, which is indicated by an  orange vertical line in each panel. This is expected, since for lower frequencies (or wavenumbers) the group velocity is smaller than the shock speed, so the wave cannot propagate ahead of the shock.

The spectra are narrow-band, but they are not consistent with a unique line, as it would be expected for cyclotron emission. This is due to the fact that the  ring-like particle distribution at the shock front possesses ultra-relativistic energies. The emission is then controlled not by the non-relativistic cyclotron maser, but rather by the ultra-relativistic  synchrotron maser instability, that generates a large number of harmonics with comparable growth rate to the fundamental \citep{hoshino_91}. 

Prominent line-like features are observed at $\sigma<1$, with the fundamental at $\omega=\omega_{\rm c, sol}$ or the second harmonic dominating the spectrum at low magnetizations (see the peak at $\omega\simeq 4\,\omega_{\rm p}$ for $\sigma=0.1$). In the transition cases with $0.1 < \sigma < 1$, we observe the generation of very strong harmonics up to $N=5$, where $N=\omega/\omega_{\rm c, sol}$, with high-order harmonics producing stronger lines than the fundamental (see the case with $\sigma=0.3$).
For $\sigma>1$ the spectrum shows much less prominent lines. As we will argue later, supplementary amplification mechanisms operate in this regime, and the characteristic frequency $\omega_{\rm c, sol}$ given by Eq.~\ref{eq:omega_peak} no longer controls the location of the spectral peak.

\begin{figure*}
	\begin{center}
		\includegraphics[width=0.95\textwidth,angle=0]{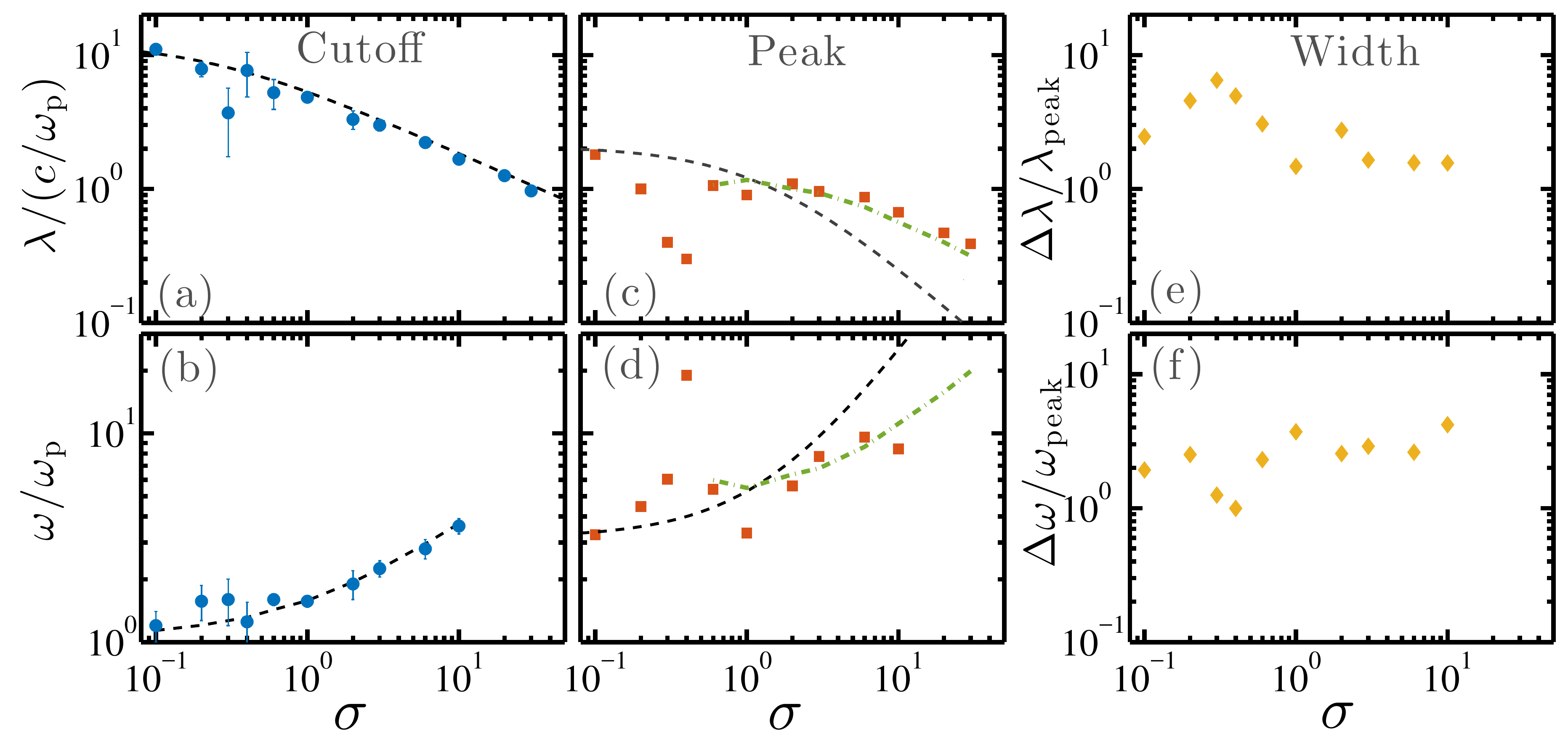}
		\caption{Characteristic wavelengths and frequencies as a function of $\sigma$. Cutoff wavelengths $\lambda_{\rm cutoff}$ and frequencies $\omega_{\rm cutoff}$ are presented in the left column (blue circles in panels a and b, respectively). Dashed black lines show the analytical predictions in Eqs.~\ref{eq:omega_cutoff} and \ref{eq:lambda_cutoff}. The wavelengths and frequencies at the peak of the spectrum ($\lambda_{\rm peak}$ and $\omega_{\rm peak}$) are plotted in the central column (panels c and d) using red squares. The black dashed lines in panels (c) and (d) show the expected $\omega_{\rm c, sol}$ from Eq.~\ref{eq:omega_peak}. The green dot-dashed line in panel (c) is the width of the density cavity at the shock front, divided by three: $L_{\rm cav}/3$ (see text, subsection~\ref{subsubsect:cavity}). The right column presents the dependence on $\sigma$ of the fractional spectral width $\Delta \lambda /\lambda_{\rm peak}$ and $\Delta \omega /\omega_{\rm peak}$ (panels e and f, respectively).}
		\label{fig:wavelength_sigma}
	\end{center}
\end{figure*}

The dependence of the relevant wavelengths and frequencies on the magnetization is presented in Figure~\ref{fig:wavelength_sigma}. The top row refers to wavelengths, the bottom row to frequencies.
The left column shows the variation with $\sigma$ of the cutoff wavelength $\lambda_{\rm cutoff}$ (panel a) and cutoff frequency $\omega_{\rm cutoff}$ (panel b). The values derived from our simulations are plotted using blue circles, and they are in very good agreement with the analytical predictions of Eqs.~\ref{eq:omega_cutoff} and \ref{eq:lambda_cutoff}, indicated by the black dashed lines. The only exception is the transition case $\sigma \simeq 0.3$,  where the low-frequency cutoff is non-stationary.

The central column (panels c and d) presents the variation with $\sigma$ of the peak wavelength and frequency (red squares are the results of our simulations), defined as the location where the precursor spectrum peaks (see Fig.~\ref{fig:spectrum_five_sigmas}). Black dashed lines indicate the expectation for soliton emission, Eq.~\ref{eq:omega_peak}. It is apparent that the peak values obtained in the simulations do not agree with the analytical estimate given by Eq.~\ref{eq:omega_peak} for any $\sigma > 0.1$.\footnote{We note, however, that in simulations with $\sigma <0.1$, not presented here, we have obtained a very good agreement between the measured $\omega_{\rm peak}$ and $\omega_{\rm c, sol}$ given in Eq.~\ref{eq:omega_peak} \citep[see also][]{gallant_92}.} 
 To understand the disagreement we define two regimes: (i) the transition cases ($0.1 < \sigma <1$) and (ii) the magnetically dominated cases ($\sigma>1$). 

In case (i), high-order harmonics in the precursor spectrum are stronger than the fundamental, and the spectral peak is not at the fundamental frequency. If we artificially select the lowest frequency corresponding to a local maximum in the spectrum, we find that its location is in reasonable agreement with the expected fundamental frequency $\omega_{\rm c, sol}$ (see  top two panels in Fig.~\ref{fig:spectrum_five_sigmas}). 
In case (ii), we do not find evidence of any strong line at the expected $\omega_{\rm c, sol}$ or its harmonics, but rather we observe less prominent lines at frequencies that have no clear connection with $\omega_{\rm c, sol}$. The measured peak frequency scales as $\omega_{\rm peak} \approx 3 \sigma^{1/2} \omega_{\rm p}$. In contrast,  from Eq.~\ref{eq:omega_peak} we would expect a stronger scaling with $\sigma$, since $\omega_{\rm c, sol} \to \sigma \omega_{\rm p}$ in the limit $\sigma \gg 1$.  As discussed in the next subsection, we attribute the observed scaling to the presence of a resonant cavity in the shock structure, that builds up only for $\sigma>1$. We show below that the peak wavelength in case (ii) corresponds to an eigenmode of the cavity, and it is roughly three times shorter than the cavity width (see the green dot-dashed lines in panels c and d). 

The right column (panels e and f, respectively) presents the dependence on $\sigma$ of the fractional spectral width in wavelength and frequency space ($\Delta \lambda /\lambda_{\rm peak}$ and $\Delta \omega /\omega_{\rm peak}$, respectively). The width $\Delta \omega$ is the difference between the two frequencies (one above the peak frequency $\omega_{\rm peak}$ and one below) where the power drops by a factor of 30 below the peak. The width $\Delta \lambda$ is defined in an analogous way.
This shows quantitatively that the spectrum is narrow, with $ \Delta \omega/\omega_{\rm peak} \lesssim 3$ nearly independently of $\sigma$. The spectra of the cases with $\sigma<1$, that show pronounced line-like features, are even narrower, with line widths of $ \Delta \omega/\omega_{\rm peak} \lesssim 1$ (see the top two panels in Figure~\ref{fig:spectrum_five_sigmas}).

\subsubsection{Resonating cavity in the shock structure at $\sigma>1$}
\label{subsubsect:cavity}
In the previous subsection we have found that  in the magnetically dominated
regime  $\sigma>1$, the peak frequency in our simulations does not scale as the expected gyration frequency in the soliton, $\omega_{\rm c,sol}$.
The physical picture that led to the estimate of $\omega_{\rm c,sol}$ must then be revised, since the shock structure 
for $\sigma>1$ appears to be different than for lower magnetizations. In fact, instead of one density peak defining the shock front, as it is the case in 
the $\sigma \lesssim 1$ regime, we observe for $\sigma \gtrsim 1$ the build-up of two density peaks separated by a cavity.\footnote{We believe that the structure of the shocks studied here is controlled by wave dispersion (rather than dissipation), given the importance of the precursor emission from the shock. For $\sigma >1$, the amount of dispersion provided by the leading soliton becomes insufficient to sustain the shock structure, and  a secondary soliton forms to provide additional dispersion.}
As we now argue, it appears that the density cavity plays an essential role in amplifying 
the precursor emission and in selecting a well-defined wavelength for the precursor waves that corresponds to an eigenmode of the cavity.

\begin{figure*}
\begin{center}
\includegraphics[width=0.97\textwidth,angle=0]{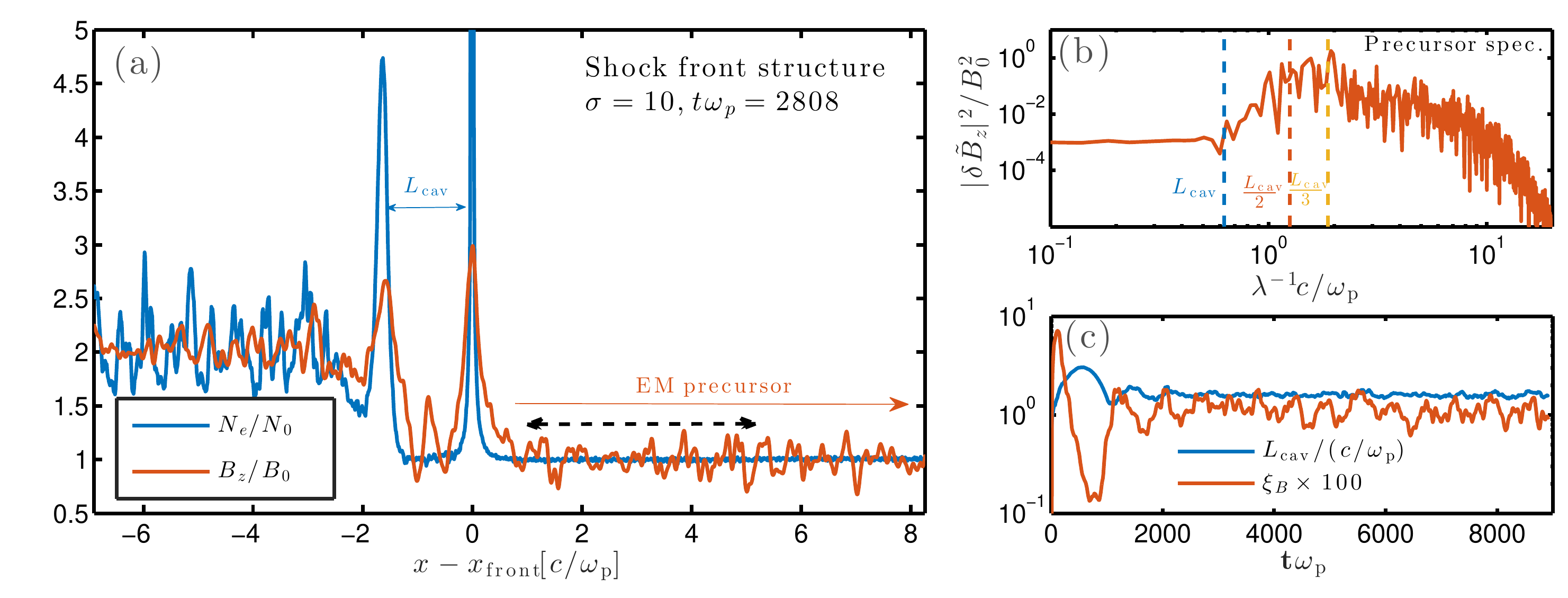}
\caption{Structure of the shock transition region for $\sigma=10$. Panel~(a): profile of the electron density $N_e/N_0$ (blue line)  and of the transverse magnetic field $B_z/B_0$ (red line). Panel (b): precursor spectrum in $k$-space,  $|\delta B_z(k)|^2/B_0^2$, as a function of the inverse wavelength $\lambda^{-1}=k/(2\pi)$. The three vertical dashed lines correspond to the cavity width $L_{\rm cav}$ (blue), to $L_{\rm cav}/2$ (orange), and to $L_{\rm cav}/3$ (red). Panel~(c): time evolution of $L_{\rm cav}$ (blue line) and of the precursor wave energy $\xi_B$  multiplied by 100 (red line). The value of $\xi_B$ was derived in the region closer to the shock front than  previously (between $1$ and $5 c / \omega_{\rm p}$ ahead of the front). The efficiency settles to a steady state at the same time as the cavity length does. }
\label{fig:cavity}
\end{center}
\end{figure*}

In Figure~\ref{fig:cavity} we illustrate the structure of the shock transition region for $\sigma=10$ 
at a well-advanced stage of the simulation when the precursor power has reached a steady 
state. Panel~(a) of this figure shows the profile of the electron density (blue line) 
and of the transverse magnetic field $B_z/B_0$ (red line). The shock front is located at 
$x-x_{\rm shock}=0$ and it propagates in the $+\hat{x}$ direction. The two density 
peaks near the shock are separated by a cavity of width 
$L_{\rm cav} \simeq 1.6\, c/\omega_{\rm p}$, just behind the shock front. The magnetic field 
profile peaks at the positions of the two density spikes, but in addition it exhibits a wave-like 
pattern within the density cavity. For this particular snapshot, only a mode with wavelength
$\lambda=L_{\rm cav}/2$ is clearly seen in the cavity. However, the cavity is dynamic in nature, and different 
eigenmodes are distinctly seen at different times. 

In panel (b) of Figure~\ref{fig:cavity} we demonstrate the role of the cavity in shaping the precursor spectrum, by showing the wavenumber spectrum 
 as a function of $\lambda^{-1}=k/(2\pi)$. 
Some characteristic emission wavelengths are easily identified. 
For instance, the cutoff wavelength at $\lambda_{\rm cutoff} \simeq 1.6\, c/\omega_{\rm p}$ 
seems to be closely related to the width of the density cavity $L_{\rm cav}$, which is indicated by a vertical dashed blue line. The other two vertical lines (red and orange, respectively) correspond 
to wavelengths equal to $L_{\rm cav}/2$ and $L_{\rm cav}/3$, respectively. The latter 
matches well the position of the strongest emission line. As discussed below, this holds for all $\sigma\gtrsim 1$.

To assess the connection between the cavity size and the precursor efficiency we show in 
 panel~(c) the time evolution of $L_{\rm cav}$ (blue line) and of the precursor wave energy $\xi_B$ 
multiplied by a factor of 100 (red line). The value of $\xi_B$ was computed in a region closer to the shock front than we have done 
before (here, between $1$ and $5\, c/\omega_{\rm p}$ ahead of the front), which allows to probe more directly the causal connection between the precursor efficiency and the instantaneous shock structure. 
This panel shows that the cavity width (blue line) initially 
increases, then it decreases and finally settles to a steady state. The time evolution of the 
precursor efficiency appears to be anti-correlated to the cavity width: when the cavity size is larger 
the emitted precursor is weaker (no amplification), and the wave intensity settles to a steady state at the same time ($\ompt\sim 1000$) as  the cavity width. We interpret this behavior as a self-regulation in the shock structure, such that the cavity width self-tunes to the value
where it can efficiently channel the precursor emission into the upstream, 
i.e., $L_{\rm cav}$ has to be roughly equal to $\lambda_{\rm cutoff}$ (see also panel b). When this condition is met, the wave is amplified and its efficiency settles to the steady state. The critical role of the cavity for efficient wave emission is also revealed by inspecting the shock profile at the time when the precursor intensity sharply increases, right before settling to a steady state ($\ompt\sim 1000$): we see that large $B_z$ fluctuations are first amplified in the cavity, and the emission of a strong precursor propagating upstream is then the consequence of partial transmission of these waves from the cavity through the leading soliton.

The validity of our ``resonating cavity'' interpretation is tested in Figure~\ref{fig:wavelength_sigma} (panels c and d), where we show that the peak wavelength of the precursor emission (red squares) is consistent with $L_{\rm cav}/3$ (green dot-dashed lines in panel c), for all  $\sigma\gtrsim 1$. In other words, for magnetically dominated plasmas
the wave amplification inside the cavity plays an important role
in selecting the dominant wavelength of the emitted precursor, as an eigenmode of the cavity.
It follows that the peak frequency for $\sigma\gtrsim 1$ scales as $\omega_{\rm peak} \simeq 3\, \omega_{\rm cutoff} \simeq 3 \sqrt{\sigma} \omega_{\rm p}\simeq 3\,\omega_{\rm c}$ in the
the simulation frame, where we have used that $\gamma_{\rm s|d}\simeq \sqrt{\sigma}$ for $\sigma\gg1$. This should be contrasted with equation~\ref{eq:omega_peak}, whose scaling ($ \propto \sigma \omega_{\rm p}$ in the $\sigma \gg 1$ limit) is not supported by our simulations.

\subsection{Wave strength parameter}
\label{subsect:strength_parameter}

The wave strength parameter (also known as ``wiggler'') measures the dynamical effect of the propagating wave on the background plasma. It is defined through the equation of motion of particles in a high-amplitude wave \citep{lyubarsky_06, iwamoto_17}:
\begin{eqnarray}
u_x &= &\gamma \beta_x = -\gamma_0 \beta_0+{a^2 \over 2} \cos^2[ \omega(x/c + t)] \, , \\
u_y &=& \gamma \beta_y =  a \cos [\omega (x/c + t)] \, .
\label{eq:quiver_uy}
\end{eqnarray}
where
\begin{equation}
a = {e\, \delta E_y \over m_e c  \omega} \,
\label{eq:wiggler_definition}
\end{equation}
is the strength parameter of the wave. $\delta E_y$ is the electric field of the wave and $\omega$ is the wave frequency; $e$ and $m_e$ are the particle charge and mass, respectively.
When $a>1$, the particle quiver motion becomes relativistic and the plasma back-reacts strongly onto the wave. We have employed two measures for the wave strength parameter: either from the maximum excursion in $u_y$, $a_{\rm max}=\max(u_y)$; or from the root mean square value, $a_{\rm std}$. These choices are motivated by the form of equation~\ref{eq:quiver_uy}, where the wiggler parameter controls the $y$-oscillations of the particle 4-velocity. In either case, we have extracted the measurement from the region  between $5$ and $105\, c/\omega_{\rm p}$ ahead of the front at the final time of the simulations.

\begin{figure}
\begin{center}
\includegraphics[width=0.48\textwidth,angle=0]{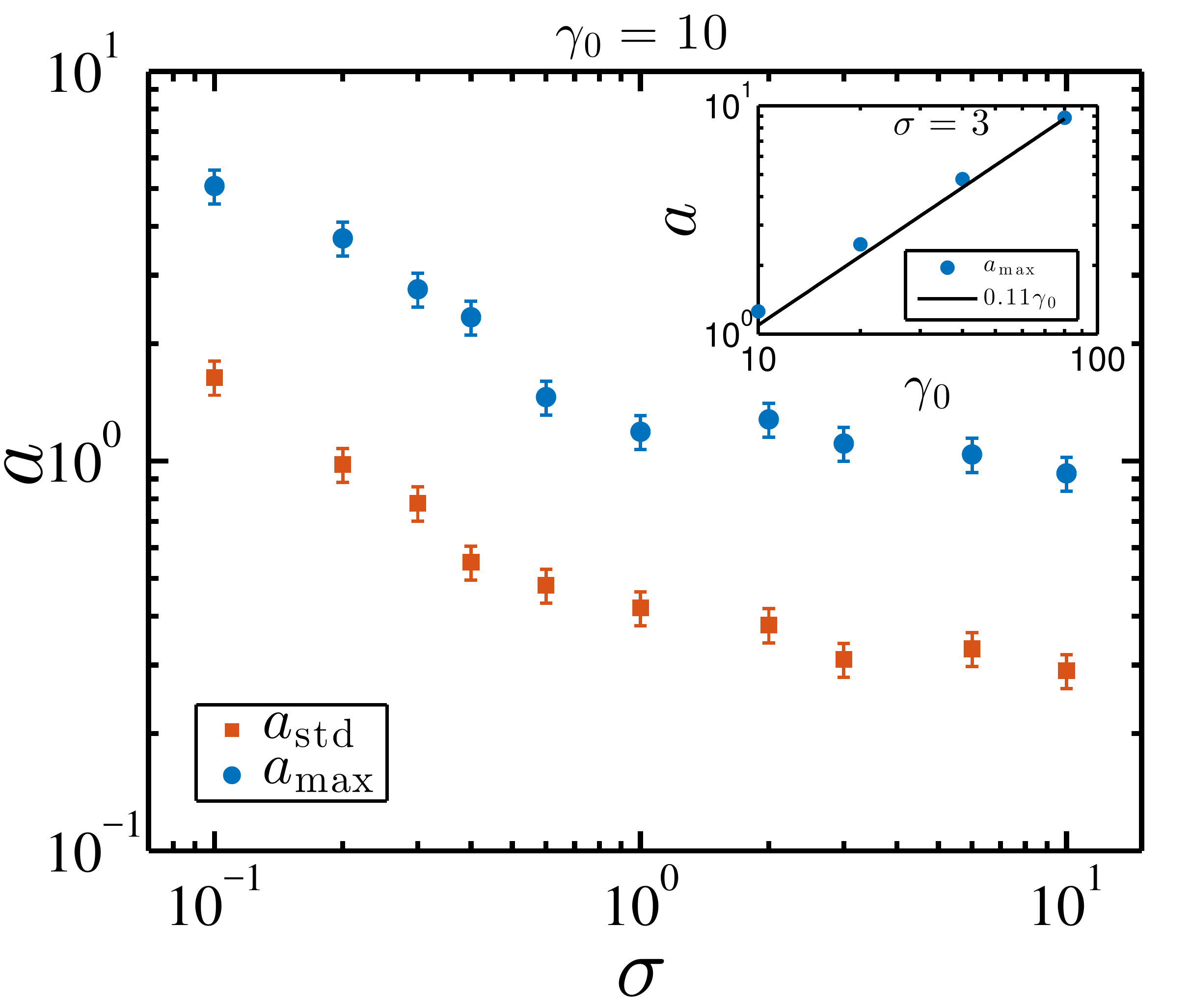}
\caption{Dependence on $\sigma$ of the wave strength parameter $a$. The two colors refer to different measurements of $a$: blue symbols correspond to the maximum value of $u_y$ in the precursor region (between $5$ and $105\, c/\omega_{\rm p}$ ahead of the front) and red symbols correspond to the root mean square value of $u_y$ in the same region. The inset presents the dependence of $a_{\rm max}$ on $\gamma_0$, for $\sigma=3$. The solid black line shows a linear scaling.}
\label{fig:wiggler}
\end{center}
\end{figure}

Figure~\ref{fig:wiggler} presents the dependence on $\sigma$ of the wave strength parameter, derived from our simulations. The maximal value $a_{\rm max}$ decreases from $\simeq 5$ for $\sigma=0.1$ down to $\simeq 1$ for $\sigma=10$, while the root mean square value has the same dependence on $\sigma$ but it is three times smaller, $a_{\rm std} \simeq a_{\rm max}/3$. The sub-panel of this figure shows the dependence on $\gamma_0$. Supplementary simulations were performed for this purpose, where we fixed $\sigma=3$. There is a clear linear dependence of $a$ on $\gamma_0$, as already suggested by \citet{iwamoto_17}. 

The linear dependence on $\gamma_0$ arises naturally from the fact that $\xi_B$ does not depend on $\gamma_0$, combined with the fact that the typical frequency of the precursor wave is $\simeq 3\, \omega_p$ for $\sigma \ll 0.1$ and $\simeq 3\,\omega_{\rm c}$ for $\sigma > 1$. It follows from Eq.~\ref{eq:wiggler_definition} that $a \approx \gamma_0 \sqrt{\xi_B \sigma}$ for $\sigma \ll 0.1$  and $a \approx \gamma_0 \sqrt{\xi_B}/3$ for $\sigma > 1$, which justifies the linear scaling with $\gamma_0$ shown in the inset of Figure~\ref{fig:wiggler}.
 
The wiggler parameter is Lorentz-invariant under transformations along the shock propagation direction, in virtue of equation~\ref{eq:quiver_uy}. Alternatively, one can note that the electric field of the wave transforms in the same way as its frequency. Values presented in figure~\ref{fig:wiggler} will then be the same in the shock rest frame and in the upstream rest frame. This is in contrast to the precursor normalized energy $\xi_B$ and the precursor spectrum, which are frame-dependent.


\section{Energetics in the shock front rest frame}
\label{sect:SRF}

The shock front rest frame (SRF) is, by definition, the frame where the shock is stationary.  In this frame the upstream plasma flows along the shock normal with a negative velocity in the $x$ direction (whose magnitude is larger than in the DRF). The downstream plasma recedes from the front along the negative $x$ direction. This frame can be naturally employed to quantify the incoming (and outgoing) momentum and energy, and so to derive the energy fraction channeled into the precursor wave.

\subsection{From the simulation frame to the shock rest frame}

So far, all the quantities related to the precursor waves have been given in the DRF, so we need to Lorentz transform them to the SRF. We will employ primed variables for the SRF.
The amplitude of the mean magnetic field transforms as
\begin{equation}
B_{0}^\prime = \gamma_{\rm s | d} \left( B_{0} + \beta_{\rm s|d} E_{0} \right) =  \gamma_{\rm s | d} B_{0} \left( 1+ \beta_{\rm s|d} \beta_0 \right) \, ,
\end{equation}
where we have used the shortcut notations $B_0=B_{z,0}$ and $E_0=E_{y,0}$. Since by transforming into the shock frame we are ``catching up'' with the precursor wave, the precursor amplitude will decrease as
\begin{equation}
\delta B_z^\prime = \gamma_{\rm s | d} \left( \delta B_z - \beta_{\rm s|d} \delta E_{y}\right) \simeq  \gamma_{\rm s | d}  \delta B_z \left( 1- \beta_{\rm s|d} \right)
\end{equation}
The $\xi_B$ parameter then transforms as \citep{gallant_92}
\begin{equation}
\xi_B^{\prime} = \xi_{B|\rm sh}={\langle \delta B_z^{\prime 2} \rangle \over B_0^{\prime 2}} = \left( {1 - \beta_{\rm s|d} \over 1+\beta_0 \beta_{\rm s|d}} \right)^2 {\langle \delta B_z^2 \rangle \over B_0^2} = \left( {1 - \beta_{\rm s|d} \over 1+\beta_0 \beta_{\rm s|d}} \right)^2 \xi_B \ .
\label{eq:transformation_xiB_sim_to_SRF}
\end{equation}

We compute directly $\xi_B^{\prime}$ with the following procedure. 
The values of $\gamma_{\rm s|d}$ and $\beta_{\rm s|d}$ obtained from our simulations (see figure~\ref{fig:xiB_sigma_3panel}, panels b and c) are used to Lorentz transform the electromagnetic fields into the SRF at a given snapshot of the simulation. Then, $\xi_B^{\prime}$ is computed directly, by averaging between $5$ and $105\, c/\omega_{\rm p}$ ahead of the front (the distance is still measured in the simulation frame).

\begin{figure}
	\begin{center}
		\includegraphics[width=0.45\textwidth,angle=0]{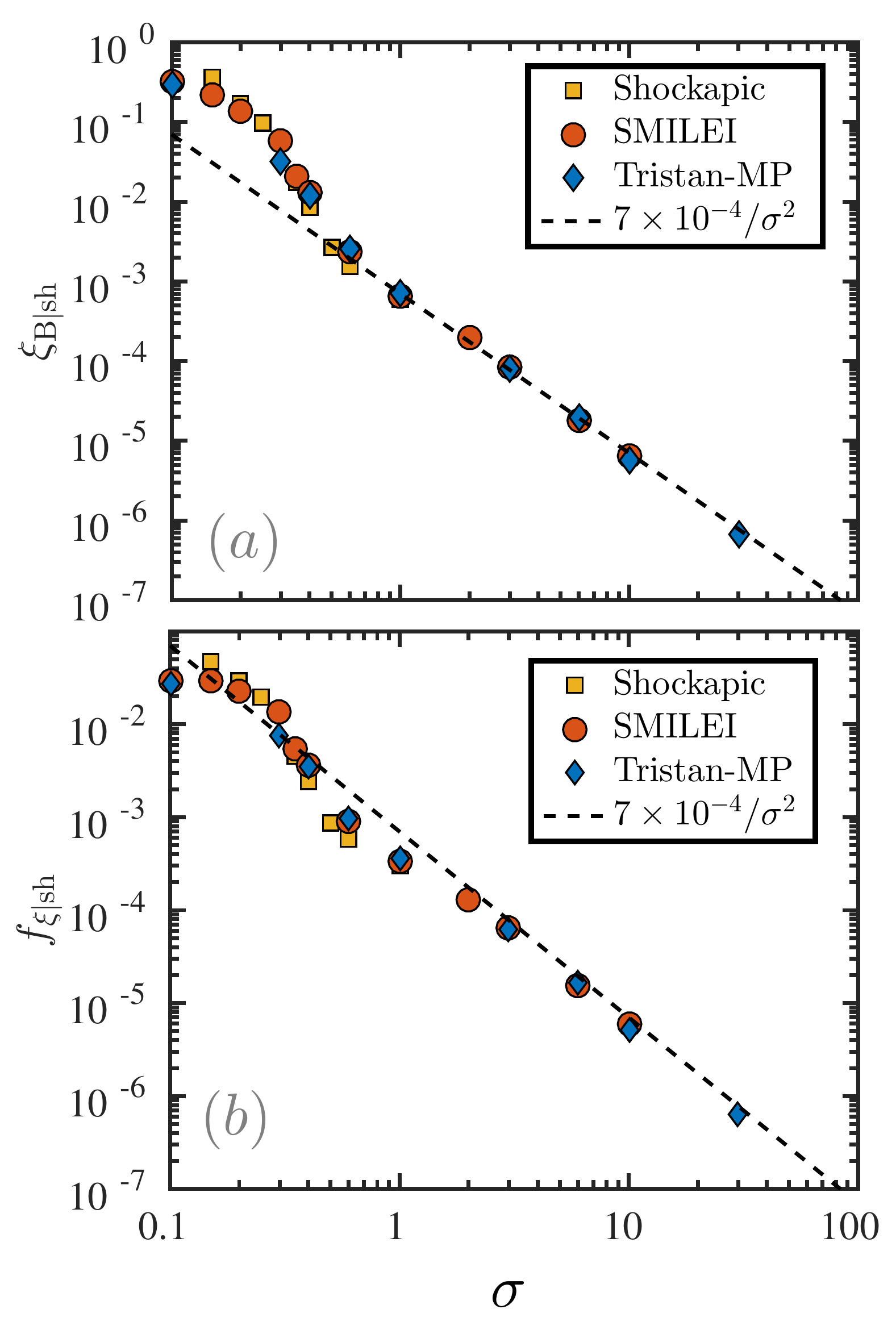}
		\caption{Energetics in the shock front rest frame: dependence on $\sigma$ of the normalized precursor wave energy $\xi_{B|\rm sh}=\xi'_B$ (top panel) and of the energy fraction in the precursor wave relative to the total incoming energy, $f_{\xi|\rm sh}=f'_\xi$ (bottom panel). Different symbols refer to a different code: red circles for \textsc{Smilei}, blue diamonds for \textsc{Tristan-MP}, and orange squares for  \textsc{Shockapic}. The dashed black lines indicate the $\propto \sigma^{-2}$ scaling.}
		\label{fig:SRF_energy_budget}
	\end{center}
\end{figure}

In Figure~\ref{fig:SRF_energy_budget} (panel a) we present the dependence of  $\xi_B^{\prime}=\xi_{B|\rm sh}$ on $\sigma$, obtained independently with the three PIC codes used in this study: orange squares for \textsc{Shockapic}, red circles for \textsc{Smilei} and blue diamonds for \textsc{Tristan-MP}. First, the figure demonstrates excellent agreement between the three codes. Second, it shows that, beyond the transition cases with $0.1<\sigma<1$, where $\xi_B^{\prime}$ attaints the largest values, the normalized wave energy in the SRF scales as $ \xi_B'\simeq7\times 10^{-4} \sigma^{-2}$
for $\sigma>1$. This scaling is plotted with  a dashed black line, and it can be easily justified. In fact, in section~\ref{sect:results} we have shown that for $\sigma \gg 1$ the wave amplitude in the DRF converges to a constant (i.e., $\sigma$-independent) value, $\xi_B \simeq 10^{-2}$. In addition, the asymptotic shock velocity in the DRF is $\beta_{\rm s | d} \simeq 1 - 1/(2\sigma)$ for  $\sigma \gg 1$. Plugging these two scalings into Eq.~\ref{eq:transformation_xiB_sim_to_SRF} leads to  $ \xi_B^{\prime} = 6.3 \times 10^{-4} \sigma^{-2}$, which is very close to the measured scaling.

\subsection{Energy budget in the precursor}

Let us also discuss the global energy budget as seen from the SRF (i.e., the fraction of total incoming energy channeled into the precursor). In the SRF, the energy conservation equation including wave contributions can be written as:
\begin{equation}
\gamma_u^{\prime 2} \beta_u^{\prime} W_u - {\delta B_{ u}^{\prime 2} \over 4 \pi} = \gamma_d^{\prime 2} \beta_d^{\prime} W_d + {\delta B_{ d}^{\prime 2} \over 4 \pi} \ .
\label{eq:Econservation_SRF}
\end{equation} 
where $W_i=w_i + b_{0,i}^2/(4\pi)$ is the generalized enthalpy, expressed in the proper frame of the fluid. The left-hand side corresponds to the upstream total energy content and the right-hand side to the downstream energy content. We have used that the upstream and downstream electromagnetic wave energies can be expressed as:
\begin{eqnarray}
{\delta B_{u}^{\prime 2} \over 4 \pi} & =& {\langle \delta B_{z,u}^{\prime 2} + \delta E_{y,u}^{\prime 2} \rangle \over 8 \pi}  \, , \\
{\delta B_{d}^{\prime 2} \over 4 \pi} & =& {\langle \delta B_{z,d}^{\prime 2} + \delta E_{y,d}^{\prime 2} \rangle \over 8 \pi}  \, , 
\end{eqnarray} 
respectively. The brackets represent either space averages at a given time or equivalently time averages at one spatial position. We have neglected the contribution from electrostatic waves, since it is largely sub-dominant in pair plasmas.

Reminding that the upstream magnetization is Lorentz invariant, we use $\sigma = \sigma^\prime = b_{0, u}^2/(4\pi w_u)$. The fraction of total incoming energy channeled into the precursor wave is then
\begin{equation}
\label{ecco}
f_{\xi}^\prime =f_{\xi |\rm sh}= {{\rm EM~wave~energy} \over {\rm Total~incoming~energy}} = {\xi_B^\prime \over \beta_{ u}^{\prime}} {\sigma \over 1+\sigma} \, ,
\label{eq:fxi_SRF}
\end{equation} 
where $\beta_{ u}^{\prime}$ is the upstream flow velocity measured in SRF. Equivalently, it is the shock front speed in the upstream rest frame. One  can also give the fraction of incoming particle kinetic energy channeled into the precursor wave:
\begin{equation}
g_{\xi}^\prime = {{\rm EM~wave~energy} \over {\rm Incoming~kinetic~energy}} = {\delta B_u^{\prime 2} \over 4 \pi \gamma_{ u}^{\prime 2} \beta_{ u}^{\prime} n_{ u} m c^2} = 
{\xi_B^\prime \over \beta_{ u}^{\prime}} {\sigma} \ .
\label{eq:precursor_over_kinetic_SRF}
\end{equation} 
In the latter equation $n_u$ is the upstream plasma proper density, including both species (so, $n_u=2 N_0/\gamma_0$). 

Getting back to the simulation results, in Figure~\ref{fig:SRF_energy_budget} (panel b) we present the dependence on $\sigma$ of the energy fraction $f_{\xi |\rm sh}=f_\xi^\prime$, as measured in the SRF. The maximum value is reached at $\sigma \sim 0.1$, where the precursor carries up to 5\% of the incoming energy. For $\sigma>0.3$ the energy content in the wave rapidly drops. Similarly to the $\xi_B^{\prime}$ scaling, there is a clear dependence as $\propto\sigma^{-2}$ for $\sigma>1$ (more precisely, $f_\xi'\simeq 7 \times 10^{-4}\sigma^{-2}$). The similarity comes from the fact that the upstream velocity is $\beta_{ u}^{\prime} \to 1$ and the upstream energy content is dominated by the magnetic field (i.e., $\sigma\gg1$). This implies from equation~\ref{ecco} that $f_\xi'\simeq \xi'_B$.  

In the limit $\sigma\gg1$, the conversion efficiency of incoming particle kinetic energy into wave energy scales as $g_\xi'\simeq \xi'_B\sigma\simeq 7 \times 10^{-4}\sigma^{-1}$. This should be contrasted with what we have obtained in the DRF, where this quantity became constant in the $\sigma \gg 1$ limit. 

We remark that the scalings reported so far have been obtained from 1D runs. While we expect that the dependence on $\sigma$ will remain unchanged in 2D and 3D, we speculate that the normalizations of $f_\xi$ and $g_\xi$ will decrease due to transverse effects that cannot be captured in 1D, e.g., wave filamentation and self-focusing through interaction with the upstream plasma.  In fact, the 2D simulations of \citet{iwamoto_17, iwamoto_18}, performed in the low magnetization regime $\sigma<0.5$, demonstrated that the wave energy is reduced typically by a factor of 3 (and up to 10), when going from 1D to 2D. However, we expect that the efficiency drop from 1D to 2D (and 3D) will be much less severe in the magnetically-dominated regime ($\sigma>1$) of interest for our work, given the rapid decrease of the wave strength parameter with magnetization (see Figure~\ref{fig:wiggler}), and so of the wave feedback onto the upstream plasma. This point will be addressed  in a forthcoming study (Sironi et al, \textit{in prep.}).

\section{Applications to FRBs}
\label{sect:discussion}


During a magnetar flare, in response to the motions of the neutron star crust, the above-lying magnetosphere is violently twisted and a strongly magnetized pulse is formed, which propagates away through the  magnetar wind. The FRB can be potentially generated at ultra-relativistic shocks resulting from the collision of the magnetized  pulse with the steady magnetar wind produced by its spin-down luminosity or by the cumulative effect of earlier flares \citep{lyubarsky_14,belo_17,waxman_17}. The train of electromagnetic waves emitted by the shock front via the synchrotron maser is the candidate FRB. Most works up to now assumed empirical values for the conversion efficiency of the shock kinetic energy into the precursor waves. These values were primarily motivated by the work of \citet{gallant_92} where, however, high-$\sigma$ simulations were not evolved long enough to reach a stationary state. Here, we use long-term simulations to quantify the steady-state energetics and spectrum of the precursor waves, for a wide range of magnetizations (up to $\sigma\gg1$). 
As we now argue, our work can provide a physically-grounded model for the origin of coherent emission in FRBs. 

First, the synchrotron maser at shocks is a coherent process, which helps explaining the extremely high brightness temperatures of FRBs. In this work, we have derived the fraction of incoming flow energy channeled into the precursor waves. If considered in the ejecta frame (post-shock frame), our simulations show that for $\sigma>1$ the emitted wave carries a fraction $f_\xi = 2 \times 10^{-3}/\sigma$ of the total energy.
This corresponds to a fraction $g_\xi \simeq 2 \times 10^{-3}$ of the incoming particle kinetic energy, regardless of $\sigma$. If one considers the energy budget in the shock rest frame, the previous scalings become $f_\xi^\prime \simeq  7 \times 10^{-4}/\sigma^2$ and $g_\xi^\prime \simeq 7 \times 10^{-4}/\sigma$, respectively.

Second, the precursor emission is linearly polarized, in agreement with the observations of several non-repeating FRBs  \citep{ravi_16, petroff_17, caleb_18} and of the repeating \frb \citep{michilli_18, gajjar_18}. Linear polarization is a natural consequence of the resonance of bunching particles with the extraordinary mode (X-mode). This mode can escape out of the plasma and become a vacuum electromagnetic wave. A contribution from the ordinary mode (O-mode) was also observed in the 2D simulations of \citet{iwamoto_18}, but it was found to be largely sub-dominant in strongly magnetized plasmas.

Third, the spectral peak can fall in the GHz range for a reasonable choice of parameters. In particular, we have found that in the post-shock frame the emission peak frequency scales as $\omega_{\rm peak} \simeq 3\, \omega_{\rm p}$ for $\sigma \lesssim 0.1$ and as $\omega_{\rm peak} \simeq 3 \,\omega_{\rm c}$ for $\sigma >1$. Several high-order harmonics characterize the transition region with $0.1 < \sigma <1$. Joining the two regimes, and neglecting for simplicity the transition cases, we can cast the peak frequency as $\omega_{\rm peak} \simeq 3\, \omega_{\rm p}\max[1,\sqrt{\sigma}]$. This can be recast in a simpler form in the pre-shock frame as
\begin{equation}
\omega^{\prime \prime}_{\rm peak} \approx 3 \gamma_{\rm s | u} \omega_{\rm p} \, ,
\end{equation}
as long as the shock is moving with an ultra-relativistic bulk Lorentz factor $\gamma_{\rm s|u}$ into the upstream medium. 
The emission frequency for an upstream observer is then
\begin{equation}
\nu'' = {\omega^{\prime \prime}\over 2\pi } \approx 2.7 \times 10^4 \gamma_{\rm s | u} \left( {n_e \over 1~{\rm cm}^{-3}}\right)^{1/2}~{\rm Hz} \, ,
\end{equation}
where $n_e$ is the pre-shock electron density. If we assume that the upstream frame corresponds to the observer frame (which is true if the pre-burst wind expands with a non-relativistic velocity), then the combination $\gamma_{\rm s | u} \sqrt{n_e / 1~{\rm cm}^{-3}} \approx 4 \times 10^4$ is required for the shock to emit in the GHz band, in rather good agreement with the estimates of \citet{belo_17}. As recently found by \citet{metzger_19} this frequency is also consistent with $\sim$GHz emission from decelerating blast waves produced by flare ejecta in young magnetars.

Finally, the spectrum is narrow-band, $\Delta \omega /\omega_{\rm peak} \lesssim 1-3$ (see figure~\ref{fig:wavelength_sigma}), which is again consistent with the observations \citep[e.g., ][]{law_17, macquart_18}.

 \subsection{Comment on criticisms to the synchrotron maser}
 A number of criticisms have recently been moved against the synchrotron maser emission as a source of the coherent FRB radiation.  \citet{lu_kumar_18} looked into a wide variety of
 maser mechanisms operating in either vacuum or plasma and found that none of them can explain
 the high luminosity of FRBs without invoking unrealistic or fine-tuned plasma conditions. Here, we argue that the synchrotron maser at relativistic shocks --- due to its unique properties --- still remains a viable candidate for powering FRBs. 
 
 First, it was argued that the synchrotron maser in vacuum requires fine-tuned plasma conditions where the magnetic field is nearly uniform (to within an angle $\gamma^{-1}$) and the
 particles' pitch-angle distribution is narrowly peaked with spread $\lesssim \gamma^{-1}$. Here, $\gamma$ is the typical Lorentz factor of the emitting particles. This is indeed the natural configuration expected at a relativistic magnetized shock, if the pre-shock particles have non-relativistic temperatures (which is anyway a requirement for efficient synchrotron maser emission). In the shock transition region, the magnetic field is nearly uniform, and the particles coherently rotate in a plane perpendicular to the field (with negligible pitch angle spread).

 Second, it was argued that it is unclear how the mechanism for the population inversion required by the maser is achieved. Once again, this is naturally realized in the shock transition of a magnetized relativistic shock, where the particles form a ring in momentum space at fixed Lorentz factor $\gamma\sim \gamma_0$, while the inner region of the ring (i.e., at lower $\gamma$) is devoid of particles, as indeed required for the existence of a population inversion.
 
 Also, it was argued that during the maser amplification process, high-energy electrons radiate faster than low-energy ones, so the population inversion condition may be quickly destroyed. This is indeed true for each generation of particles passing through the shock, since the synchrotron maser instability relaxes by ``filling up'' the hollow ring in momentum space, thus destroying the population inversion. However, while this happens, a new generation of particles is entering into the shock. They establish a new ring in momentum space, and keep sustaining the radiated train of precursor waves. In other words, the continuous passage of plasma through the shock ensures that the population inversion is steadily maintained (yet, at each time by different particles).
 
Finally, \citet{lu_kumar_18} considered more specifically the maser synchrotron emission at shocks, which they named as ``bunching in the gyration phase.'' In order to minimize the effect of induced Compton scattering, they estimated that the radiative efficiency of the shock must be extremely small. However, they considered only internal shocks occurring in between two identical consecutive density shells propagating inside the pre-burst wind, and not  the leading shock moving directly into the  wind. Aside from the limitations of induced Compton scattering, it is anyway hard for internal shocks to be efficient emitters of maser synchrotron radiation, since they propagate
into a relativistically hot shocked plasma (the downstream region of the leading shock). The arguments by \citet{lu_kumar_18} will not apply to the leading shock. First, this shock is likely to be ultra-relativistic, unlike internal shocks. Second, the properties of the shell and of the pre-burst wind (as regard to magnetization, temperature, composition) are generally different, in contrast to what \citet{lu_kumar_18} implicitly assumed. We believe that the quantitative results on precursor energetics and spectrum that we provide in this work will help revisit the estimates provided by \citet{lu_kumar_18}, for the case of the leading shock.

\section{Summary and conclusions}
\label{sect:conclusion}
In this work we have investigated by means of 1D Particle-In-Cell simulations the physics of synchrotron maser emission from perpendicular relativistic shocks that propagate in highly magnetized electron-positron plasmas (with magnetization $0.1 \leq \sigma \leq 30$). For strongly magnetized shocks, we expect that multi-dimensional simulations (to be discussed in a forthcoming work) will not yield very different results than what we present here. 
We have explored the efficiency and spectrum of the electromagnetic precursor emission as a function of $\sigma$ and $\gamma_0$. We have found that:
\begin{enumerate}
\item The shock front emits efficiently and steadily a train of high-amplitude electromagnetic precursor waves for any $\sigma$ and $\gamma_0$, in the range $0.1 \leq \sigma \leq 30$ and $\gamma_0\geq 5$ that we have explored. The emission is linearly polarized, with fluctuating magnetic field along the same direction as the upstream mean field. 
\item Thanks to unprecedentedly long simulations, we have been able to reach the stage when the precursor emission settles to a steady state, which allows to systematically extract the wave properties (energetics and spectrum). We find that the ratio of the wave energy to the upstream magnetic energy, $\xi_B$, decreases rapidly from $3.5$ at $\sigma=0.1$ down to $0.04$ at $\sigma=1$, as measured in the post-shock frame of the simulations. For $\sigma \gg 1$, this ratio converges to a constant value $\xi_B \simeq 0.01$.  In the shock rest frame, the asymptotic scaling in the limit $\sigma\gg1$ becomes $\xi_B^\prime \propto \sigma^{-2}$.
\item For $\sigma>1$, the energy output in precursor waves normalized to the total incoming energy scales as $f_\xi \simeq 2\times 10^{-3} \sigma^{-1}$ in the post-shock frame and as $f_\xi^\prime \simeq 7\times 10^{-4} \sigma^{-2}$ in the shock rest frame. The former implies that in the downstream frame,  $\sigma>1$ shocks convert a constant fraction of the incoming particle kinetic energy into precursor waves (equal to $g_\xi\simeq 2\times 10^{-3}$).
\item Magnetically dominated shocks with $\sigma >1$ exhibit a resonating cavity in the shock front structure in between two solitons, instead of the single soliton loop that is observed for $\sigma \ll 1$ shocks. This cavity plays an essential role in amplifying the radiation and selecting the dominant emission frequency as an eigenmode of the cavity. This effect causes the peak emission frequency, as measured in the downstream frame, to scale as $ \omega_{\rm peak}\simeq 3\, \omega_{\rm c}=3\sqrt{\sigma}\omega_{\rm p}$ for $\sigma>1$, whereas earlier works  \citep{gallant_92}  quote a stronger scaling with magnetization, $ \omega_{\rm peak}\simeq\sigma\omega_{\rm p}$. 
\item The characteristic frequency of the emission, as measured in the post-shock frame, is $\omega \simeq 3 \omega_{\rm p}$ for weakly magnetized shocks $\sigma \leq 0.1$, and $\omega \simeq 3 \omega_{\rm c}$ for $\sigma \gg 1$, as we have just discussed. In the transition region $0.1<\sigma < 1$, prominent high-order harmonics of $\omega_{\rm c,sol}$ (given in Eq.~\ref{eq:omega_peak}) were observed along with the fundametal at $\omega_{\rm c,sol}$. Aside from the transition cases, we can interpolate between the low- and high-magnetization results and state that the peak emission occurs at $\omega_{\rm peak} \simeq 3\, \omega_{\rm p}\max[1,\sqrt{\sigma}]$, as measured in the downstream frame. In the pre-shock frame (which coincides with the observer frame, if the magnetar wind is non-relativistic),  this can be recast in a simpler form as $\omega^{\prime \prime}_{\rm peak} \approx 3 \gamma_{\rm s | u} \omega_{\rm p}$, where $\gamma_{\rm s|u}$ is the shock Lorentz factor in the upstream frame.
\item The spectrum of the precursor is narrow-band,  $\Delta \omega /\omega_{\rm peak} \lesssim 1-3$, with a low-frequency cutoff at $\omega_{\rm cutoff} = \gamma_{\rm s | d} \omega_{\rm p}$ (here, $\gamma_{\rm s | d}$ is the shock Lorentz factor in the downstream frame) set by the requirement that the group velocity be faster than the shock speed. 
\item We did not observe any dependence on $\gamma_0$ of the energy fraction, $\xi_B$, and of the characteristic emission frequency, $\omega_{\rm peak}/\omega_{\rm p}$, in the post-shock frame.
\end{enumerate}

We conclude with a few caveats. First, we have assumed that the upstream plasma has negligible thermal spread, $k_{B}T_0/m_e c^2=10^{-4}$. Higher temperatures are likely to suppress high-order harmonics  and reduce the global energy of the wave. Second, we have mostly focused on strongly magnetized ($\sigma>1$) plasmas, a regime that so far has received little attention. Even though this work only presents 1D simulations, we anticipate that the multi-dimensional physics of $\sigma>1$ shocks  (Sironi et al., \emph{in prep.}) will not depart significantly from what we report here. In contrast, for weaker magnetizations ($\sigma \ll 10^{-2}$),  transverse effects (e.g., Weibel-driven filamentation) will significantly reduce the energy carried by the precursor waves \citep{sironi_13, iwamoto_17}. In summary, both higher pre-shock temperatures and multi-dimensional effects at low $\sigma$ are expected to degrade the precursor efficiency, which might become too low to explain the FRB emission.
 
Finally, we have only considered electron-positron shocks. Recently, 
a very large Faraday Rotation Measure (RM) of $\sim 10^5$~rad\,m$^{-2}$  was reported from the repeating \frb  \citep{michilli_18}. This challenges the pure electron-positron composition assumed in this study, since the presence of an appreciable fraction of ions is required to produce non-zero RM \citep{margalit_18b}. This urges to explore the shock physics for electron-proton and electron-positron-proton compositions. Yet, it is still possible that the FRB pulse is produced in localized regions with pristine electron-positron composition, even though most of the magnetar wind (which inflates the surrounding nebula, where the RM accumulates) is proton-dominated.

\section*{Acknowledgments}
IP acknowledges discussions with Anatoly Spitkovsky, Patrick Crumley and Yuri Cavecchi. LS is grateful to Brian Metzger for many inspiring discussions. IP was supported by NSF grants PHY-1804048 and PHY-1523261. This work was facilitated by the Max-Planck/Princeton Center for Plasma Physics. LS acknowledges support from NASA ATP 80NSSC18K1104. The simulations were performed on Habanero cluster at Columbia University, NERSC (Edison) and NASA (Pleiades) resources, PICSciE-OIT High Performance Computing Center and Visualization Laboratory at Princeton University, and on CALMIP supercomputing resources at Universit\'e de Toulouse (France) under the allocation 2016-p1504. 

\bibliographystyle{mnras}
\bibliography{SMshock_1D}

%

\appendix
\section{Codes comparison}
\label{sect:appendix_codes}

\begin{table*}
\begin{center}
\begin{tabular}{|c|c|c|c|c|c|c|c|c|}
\hline 
PIC code & $\Delta t \,\omega_{\rm p}$ & $T_{\rm sim} \omega_{\rm p}$ & $\Delta x/(c/\omega_{\rm p})$ & $N_{\rm ppc}$  & $k_BT_0/(m_e c^2)$ & $\sigma_{\rm min}$ & $\sigma_{\rm max}$ & $\gamma_0$ \\ 
\hline 
\textsc{Tristan-MP} & 1/200 & $\lesssim 2\times10^4$ & 1/100 & 64 & $10^{-4}$ & $10^{-1}$ & 30 & 10 \\ 
\textsc{Smilei}  & $1/224$ \& $1/113$ & $6.7\times10^3$  & 1/112 & 20 & $10^{-4}$ & $10^{-1}$ & 30 & 10  \\ 
\textsc{Shockapic}  & 1/90 & $1.2\times10^3$ & 1/44.7 & 20 & $10^{-4}$ & $10^{-3}$ & 1 & 10  \\ 
\textsc{Smilei} (2)  & $1/90$ & $1\times10^3$  & 1/44.7 & 20 & $10^{-6}$ & $10^{-3}$ & 2 & 10  \\ 
\textsc{Smilei} (3)  & $1/90$ & $1.5\times10^3$  & 1/44.7 & 20 & $10^{-4}$ & $10^{-3}$ & 2 & 160  \\ 
\hline 
\end{tabular} 
\end{center}
\caption[]{Typical parameters of the PIC simulations presented in this study: $\Delta t$ is the time-step in units of the inverse plasma frequency $\omega_{\rm p}^{-1}$ (defined with both species),  $T_{\rm sim}$ is the simulation timespan, $\Delta x$ is the cell size in units of $c/\omega_{\rm p}$, $k_BT_0$ is the upstream thermal energy in units of $m_e c^2$, $N_{\rm ppc}$ is the number of particles-per-cell for each species, $\sigma_{\rm min}$ and $\sigma_{\rm max}$ are the minimal and maximal values of the magnetization explored with a given code.}
\label{table:sim_params}
\end{table*}

In this appendix we show how the results from the different codes compare.
The  synchrotron maser emission occurs through the resonance of the cyclotron harmonics
with the X-mode branch. It is not guaranteed that a typical PIC code can capture accurately
a large number of these resonances, especially at the high-frequency end of the branch. 
For instance, in typical Yee-type second order solvers of Maxwell's equations the numerical speed of electromagnetic waves is known to be artificially suppressed at high $\omega$ if the CFL number 
is smaller than unity \citep{birdsall}.

For this reason we undertook an extensive comparison of three different PIC codes: two Finite-Difference Time-Domain (FDTD) codes, 
\textsc{Tristan-MP} and \textsc{Smilei}, and the pseudo-spectral code \textsc{Shockapic}. In principle, 
\textsc{Shockapic} is the best suited to capture the dispersion relation of waves in 
a plasma, but it is the least optimized among the three codes, making it challenging 
to perform long-term simulations. Concerning \textsc{Smilei}, it is a well-optimized code, but in 1D setups it currently has only a 
standard Yee solver.  \textsc{Tristan-MP} is the most optimized for shock 
setups. Also, it allows to use a fourth-order scheme to solve Maxwell's equations \citep{greenwood_04} that 
reproduces accurately the dispersion relation of electromagnetic waves even at low CFL numbers. This is the reason why the runs presented in the main body of the paper were performed with  \textsc{Tristan-MP}.  In general, each code employs different algorithms and 
implementations. The agreement between the three codes will then be a strong indication of the physical 
robustness of our results.  

The simulation parameters for each code are presented in 
Table~\ref{table:sim_params}. The table reports the space and time resolution, the simulation timespan, the number of particles per cell,
the values of the upstream temperature $T_0$ and bulk Lorentz factor $\gamma_0$, and the explored range of $\sigma$. For better comparison we used comparable space and time resolutions: the skin depth was resolved with 100 cells in \textsc{Tristan-MP}
simulations, with 112 cells in \textsc{Smilei} simulations, and with 44.7 cells in \textsc{Shockapic} simulations.
The latter has a twice smaller resolution due to code performance limitations (not parallelized).
We noticed that a resolution lower than 20 cells per skin depth affected negatively the results
for any $\sigma$. The results become stable for any resolution higher than 40 cells per $c/\omega_{\rm p}$, 
as long as $\sigma\leq10$.
Similar conclusions were reached by \citet{iwamoto_17}.
For this reason a high spatial and time resolution was employed in the simulations presented in the main body of the paper.
Only short simulations were affordable with  \textsc{Shockapic}. For this reason, the $\sigma>1$
regime was not explored with this code (as we have discussed, at high $\sigma$ it takes longer to reach a steady state). With \textsc{Tristan-MP} and \textsc{Smilei} it
was possible to reach the stationary state for $\sigma$ up to 30. 
Concerning the number of particles per cell, the results are very weakly dependent on $N_{\rm ppc}$,
as long as at least a dozen of particles per cell are initialized.

In the following we present in more detail the comparison of precursor energy and spectrum
as derived from different codes.

\subsection{Precursor energy}

\begin{figure}
	\begin{center}
		\includegraphics[width=0.48\textwidth,angle=0]{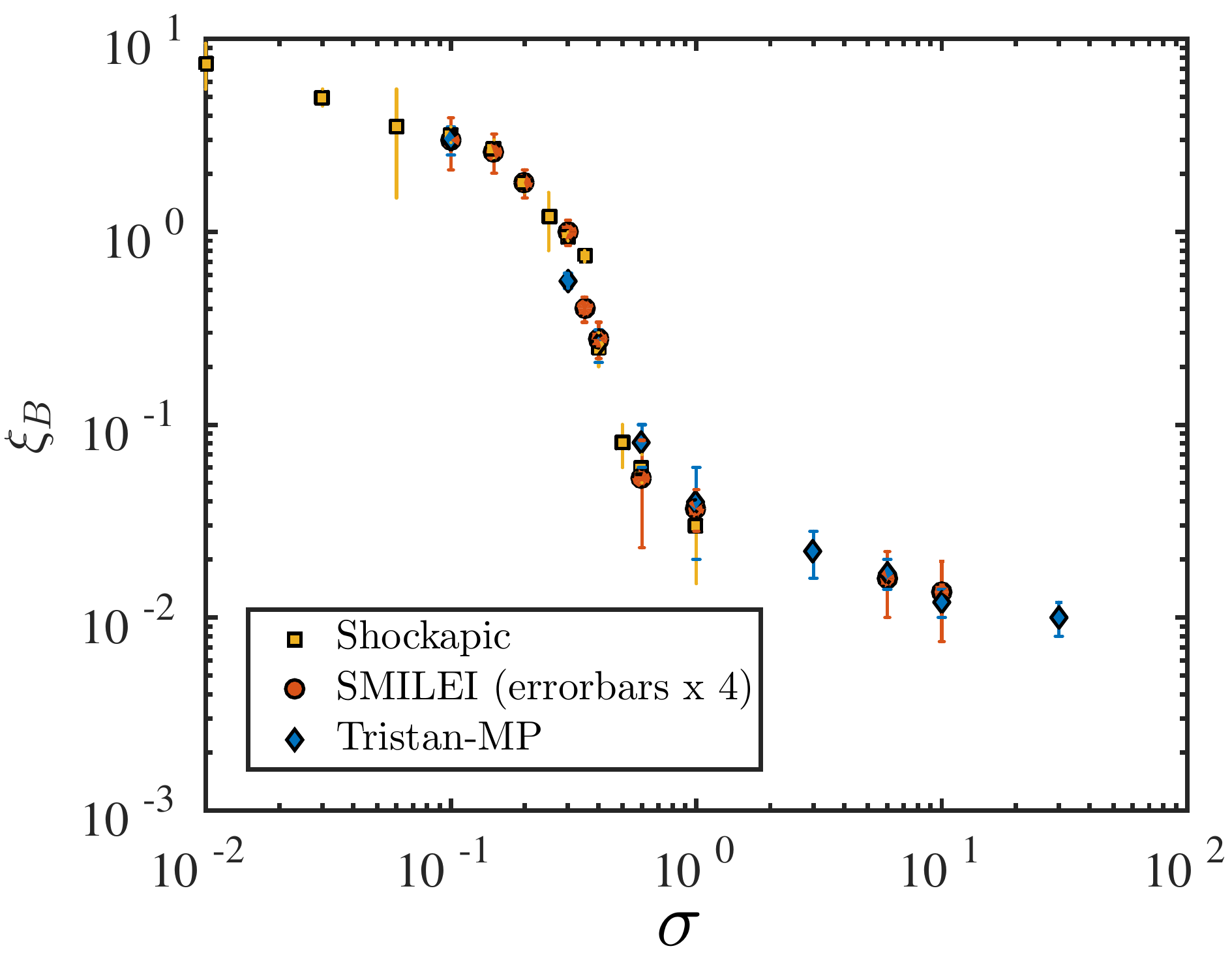}
		\caption{ Precursor wave energy $\xi_B$ as a function of $\sigma$: 
			comparison between three PIC codes used in the present study (\textsc{Tristan-MP}, \textsc{Smilei}, and \textsc{Shockapic}). Values obtained with \textsc{Shockapic}, \textsc{Smilei}, and \textsc{Tristan-MP} are plotted using orange squares, red circles, and blue diamonds, respectively. }
		\label{fig:codes_compar_xiB}
	\end{center}
\end{figure}

\begin{figure}
	\begin{center}
		\includegraphics[width=0.48\textwidth,angle=0]{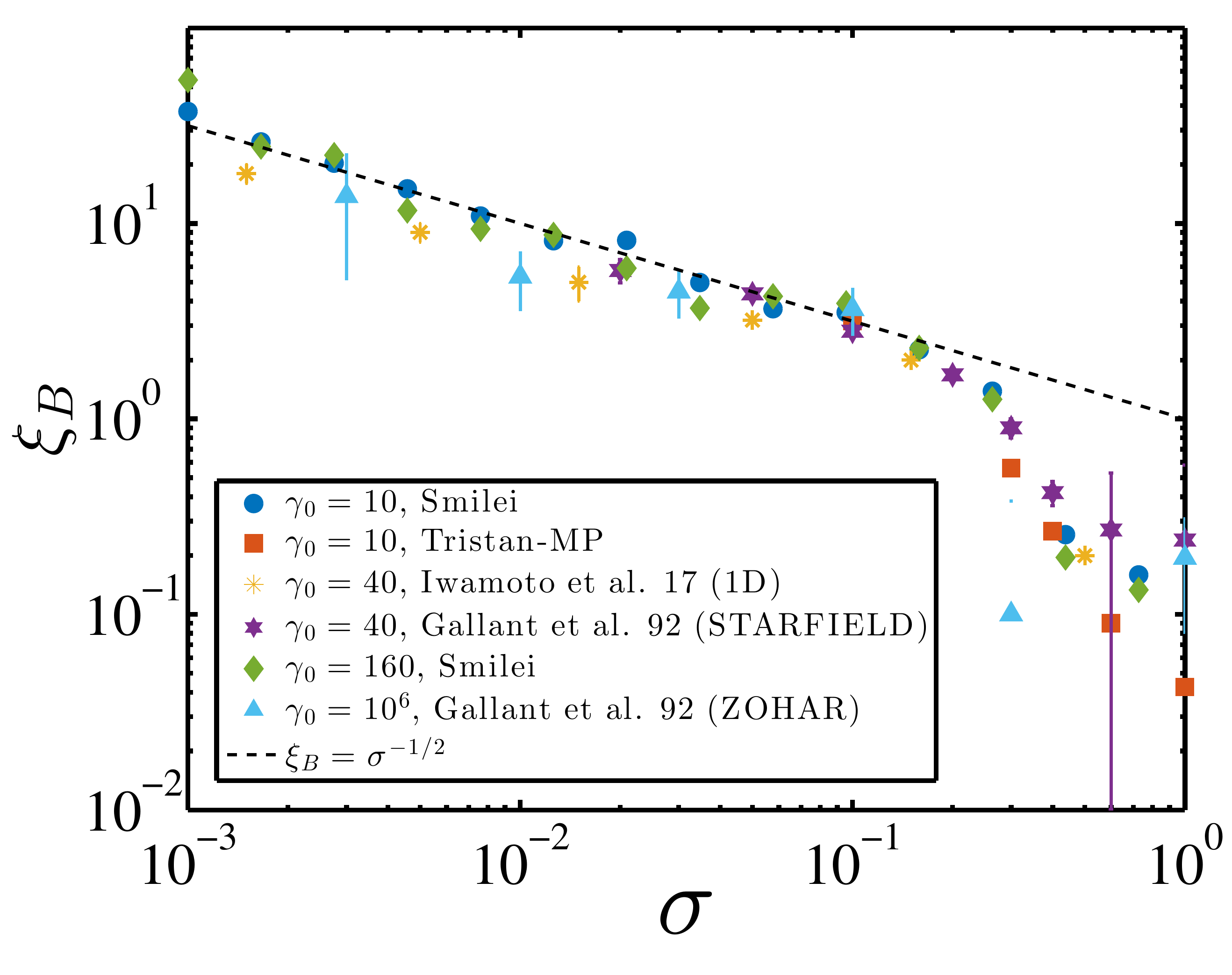}
		\caption{Same as Figure~\ref{fig:codes_compar_xiB}, but comparing the results from PIC codes in this work with earlier studies from the literature. The explored values of $\sigma$  range here from $10^{-3}$ to $1$, since other studies did not explore highly magnetized cases with sufficiently long simulations. Also, results with different values of $\gamma_0$ are shown here.  Blue circles correspond to \textsc{Smilei} simulations with $\gamma_0=10$,  red squares are from \textsc{Tristan-MP} with $\gamma_0=10$, orange stars present the data taken from \citet{iwamoto_17} with $\gamma_0=40$ (1D), magenta stars are from \citet{gallant_92} with $\gamma_0=40$, green diamonds present \textsc{Smilei} results with $\gamma_0=160$, and the values presented using light blue triangles are taken from \citet{gallant_92} with $\gamma_0=10^6$. The dashed black line presents the scaling $\xi_B=1/\sqrt{\sigma}$, which roughly fits the data points in the range $\sigma \in [10^{-3},0.1]$.} 
		\label{fig:codes_compar_xiB_litterature}
	\end{center}
\end{figure}

In figure~\ref{fig:codes_compar_xiB} we present the normalized wave energy $\xi_B$ as a function of $\sigma$, obtained with the three codes. Values obtained with \textsc{Shockapic}, \textsc{Smilei}, and \textsc{Tristan-MP} are plotted using orange squares, red circles, and blue diamonds, respectively.  In the overlapping range of $\sigma$,  we observe good agreement among different codes. For instance, in the $\sigma>1$ regime \textsc{Smilei} and \textsc{Tristan-MP} give the same values of $\xi_B$. In the range $0.1<\sigma \leq 1$, where all codes overlap, the scatter among codes is slightly larger, although the rapid drop in $\xi_B$ is common to all codes, and it happens around the same $\sigma$. We note that the transition is more abrupt in  \textsc{Shockapic}  than in \textsc{Smilei} and \textsc{Tristan-MP}, but differences remain minor.

In figure~\ref{fig:codes_compar_xiB_litterature} we extend the comparison to different studies in the literature and to different values of $\gamma_0$, from $10$ to $10^6$. The range of $\sigma$ in this figure is from $10^{-3}$ to $1$, since other studies did not explore highly magnetized cases with sufficiently long  simulations (i.e., they did not reach a steady state in the regime $\sigma\gg1$). Blue circles and green diamonds report the values obtained with \textsc{Smilei} using $\gamma_0=10$ and $160$, respectively. Both give nearly the same values for any explored $\sigma$, confirming that $\xi_B$ does not depend on the flow Lorentz factor.
 Red squares report the values from  \textsc{Tristan-MP} using $\gamma_0=10$ (same as in figure~\ref{fig:codes_compar_xiB}). The data from the 1D simulations of \citet{iwamoto_17} using $\gamma_0=40$ are plotted with orange stars. Their values are slightly smaller than what is found in this study, though generally in good agreement. Violet stars and light-blue triangles report the values from \citet{gallant_92} using $\gamma_0=40$ and $10^6$, respectively. We notice that all codes provide the same results in the range of $\sigma \in [10^{-3},0.3]$, regardless of $\gamma_0$. 
This demonstrates that the precursor wave normalized energy $\xi_B$ is not dependent on $\gamma_0$, and that our study is in very good agreement with earlier results.

 For $\sigma>0.3$ there is a noticeable scatter between different simulations.
 The most plausible reason for the discrepancy among different datasets is that the high-$\sigma$ simulations from earlier studies were not evolved long enough to reach the asymptotic state, so the value of $\xi_B$ was not yet stabilized (see, Fig.~\ref{fig:xiB_time_evolution} for the time convergence of the efficiency).

\subsection{Precursor spectrum}

\begin{figure}
	\begin{center}
		\includegraphics[width=0.48\textwidth,angle=0]{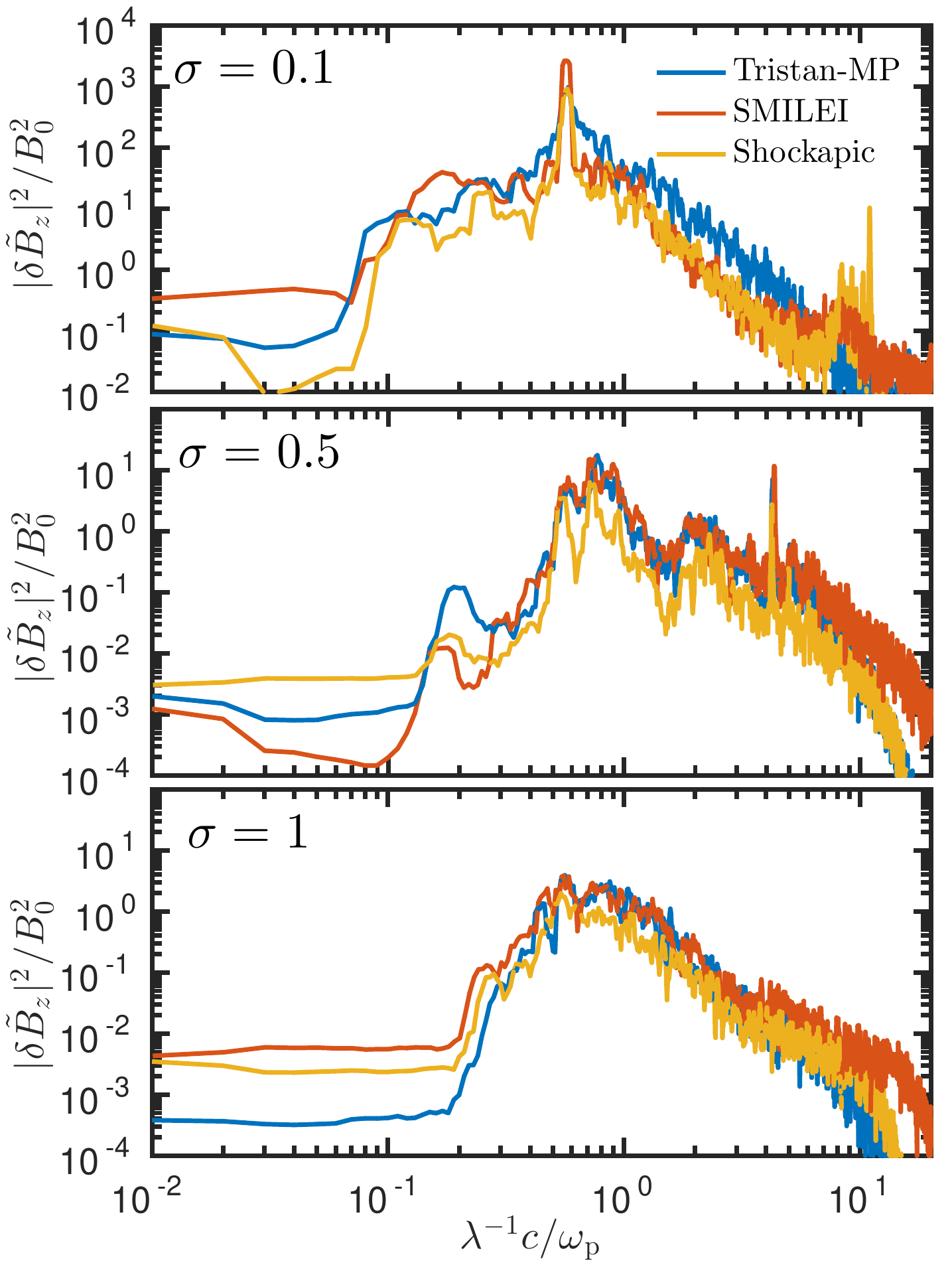}
		\caption{Precursor spectrum in $k$space for three values of $\sigma$, as indicated in the legend.  We perform a comparison among the three PIC codes used in this study: the spectrum extracted from \textsc{Tristan-MP} simulations is plotted using a solid blue line, whereas red lines are used for  \textsc{Smilei} and orange lines for \textsc{Shockapic}.}
		\label{fig:codes_compar_spectrum}
	\end{center}
\end{figure}

We now compare the precursor  $k$-spectrum among the three codes. Some differences are expected, since the numerical schemes for the integration of Maxwell's equations differ among the codes. 

In figure~\ref{fig:codes_compar_spectrum} we compare the precursor spectrum extracted from the three codes for a few representative  values of magnetization. From top to bottom, the value of $\sigma$ is $0.1$, $0.5$ and $1$, respectively. We cannot perform any comparison for $\sigma>1$ as this range was not explored with \textsc{Shockapic} (but see below, for a comparison between  \textsc{Smilei} and \textsc{Tristan-MP} at $\sigma=30$). The spectrum extracted from \textsc{Tristan-MP} is plotted using a solid blue line. Red and orange lines are used for \textsc{Smilei} and \textsc{Shockapic}, respectively. There is generally a good agreement among the codes for all values of  $\sigma$ as regard to the low-$k$ cutoff wavenumber, the high-$k$ slope, and the main peaks in the spectrum. For example, the dominant emission line for $\sigma=0.1$ and the high-order harmonic line at $\lambda=0.24\, c/\omega_{\rm p}$ for $\sigma=0.5$  are exactly at the same wavelength for the three codes. One difference can be noted: the spectral energy density is slightly smaller in \textsc{Shockapic}  than in the two FDTD codes around $\lambda^{-1}c/\omega_{\rm p} \sim 1$, for $\sigma=0.5$ and $\sigma=1$. Yet, this difference is not systematic and the overall energy in the precursor is very close among the three codes.

\begin{figure}
\begin{center}
\includegraphics[width=0.48\textwidth,angle=0]{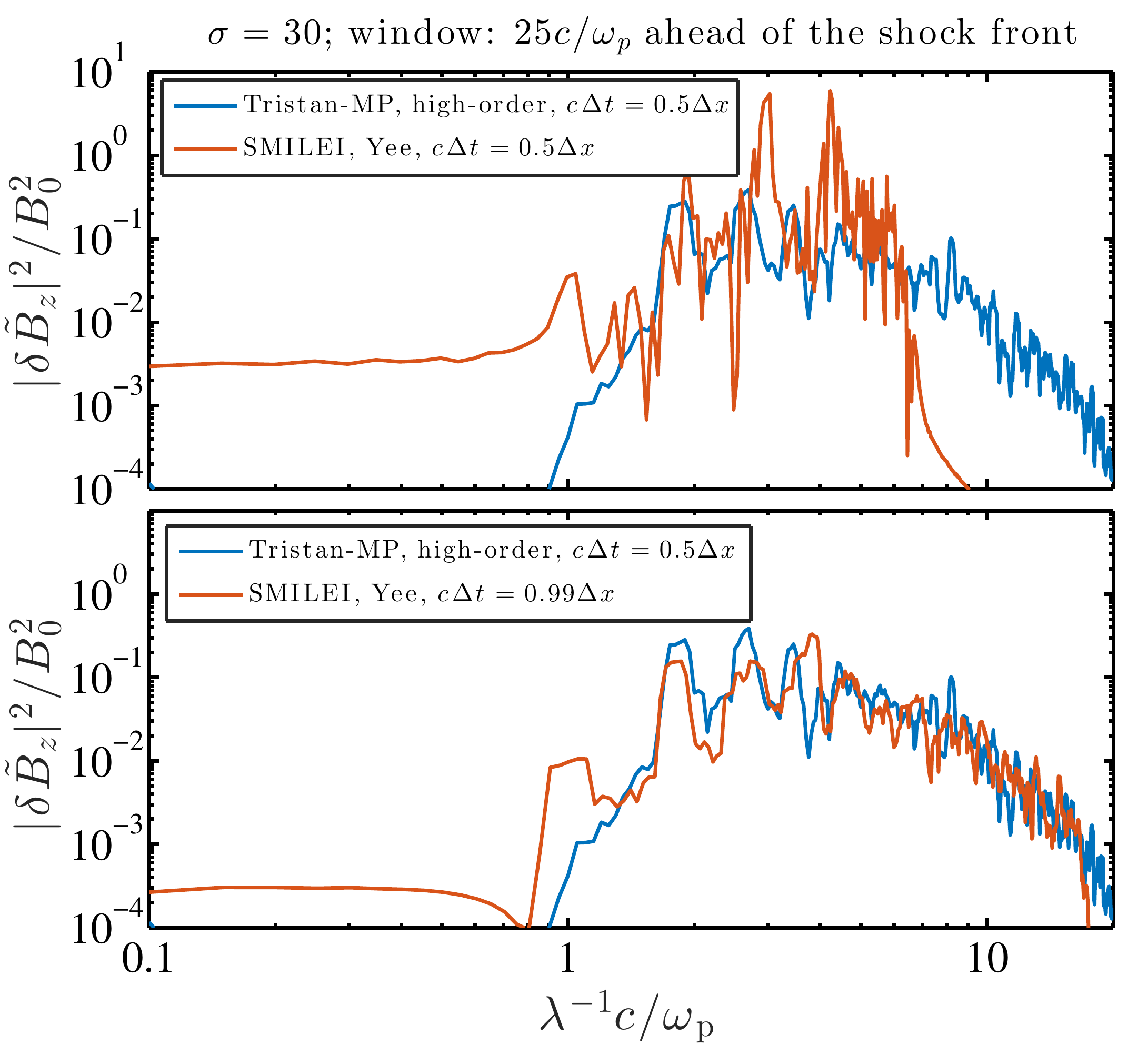}
\caption{Comparison of the precursor spectrum in $k$-space obtained from \textsc{Tristan-MP} and \textsc{Smilei} for $\sigma=30$ (the largest magnetization that we have explored, where differences among codes are most dramatic). 
The upper panel presents the spectrum from \textsc{Tristan-MP} using a fourth-order scheme to solve Maxwell's equations (blue line) and from \textsc{Smilei} using a Yee-type scheme with CFL number$=c\Delta t/\Delta x=0.5$ (red line). The latter presents a sharp cutoff at high-$k$ (i.e., for $\lambda < 0.14 \,c/\omega_{\rm p}$) and irregular line-like emission features. The lower panel presents the same comparison but with CFL=0.99 for \textsc{Smilei} (red line). The high-$k$ cutoff disappears and a very good agreement with \textsc{Tristan-MP} is obtained. We note that the spectra presented in this figure are normalized to unity, instead of the previously adopted normalization $\int |\delta \tilde{B}_z(k)|^2 /B_0^2 {\rm d}k  =\int |\delta \tilde{B}_z(\omega)|^2 /B_0^2 {\rm d}\omega = \xi_B$.}
\label{fig:codes_compar_spectrum}
\end{center}
\end{figure}

As an exception and a word of caution, 
we noticed that the use of a small CFL number with a Yee-type solver of Maxwell's equations (as used in the \textsc{Smilei} code) has a negative 
impact on the results for the largest magnetizations explored here, i.e., $\sigma>10$. 
In fact, the emission peaks at high frequencies where the light-wave branch is 
affected by the artificial reduction of the phase speed. The spectrum of the precursor 
is then sharply cut at high frequencies, affecting the overall energy 
output in the precursor.  This effect is evidenced in figure~\ref{fig:codes_compar_spectrum} for $\sigma=30$ (the largest value explored in this work).
The upper panel of the figure compares the spectrum from  \textsc{Tristan-MP} (blue), where a fourth-order scheme was used, with
the spectrum from \textsc{Smilei} (red), which employs a Yee-type scheme with $c\Delta t/\Delta x=0.5$. There is an artificial suppression in the high-$k$ region
in the \textsc{Smilei} simulation. The bottom panel shows the same comparison, but with $c\Delta t/\Delta x=0.99$ being used with \textsc{Smilei}.
In this case, the spectra agree very well, up to details in line-like features. This conveys that the high-$k$ (and so, high-$\omega$) part of the precursor spectrum can be properly captured only when the numerical integrator is capable of reproducing correctly the dispersion relation of electromagnetic waves.
This problem does not arise in \textsc{Tristan-MP} (with high-order spatial solver) and \textsc{Shockapic}, since for them the numerical dispersion of the light-wave branch is much closer to the realistic one even for small CFL numbers.

All \textsc{Smilei} simulations that use a CFL number as close as possible 
to unity (CFL=0.99) display spectra that are in very good agreement with the  other two codes for any $\sigma$.

\subsection{Concluding remark}
We find that our results do not depend on the code that we employ if these three conditions are realized: (i) a high spatial resolution (i.e., large $c/\omega_{\rm p}$) is employed; (ii) in   a Yee-type based code, the CFL number is as close as possible to unity; (iii) the simulations are sufficiently long to reach the steady state.

\section{Precursor energetics: 1D vs multi-dimensional simulations}
\label{sect:appendix_1D_vs_2D3D}

\begin{figure}
	\begin{center}
		\includegraphics[width=0.48\textwidth,angle=0]{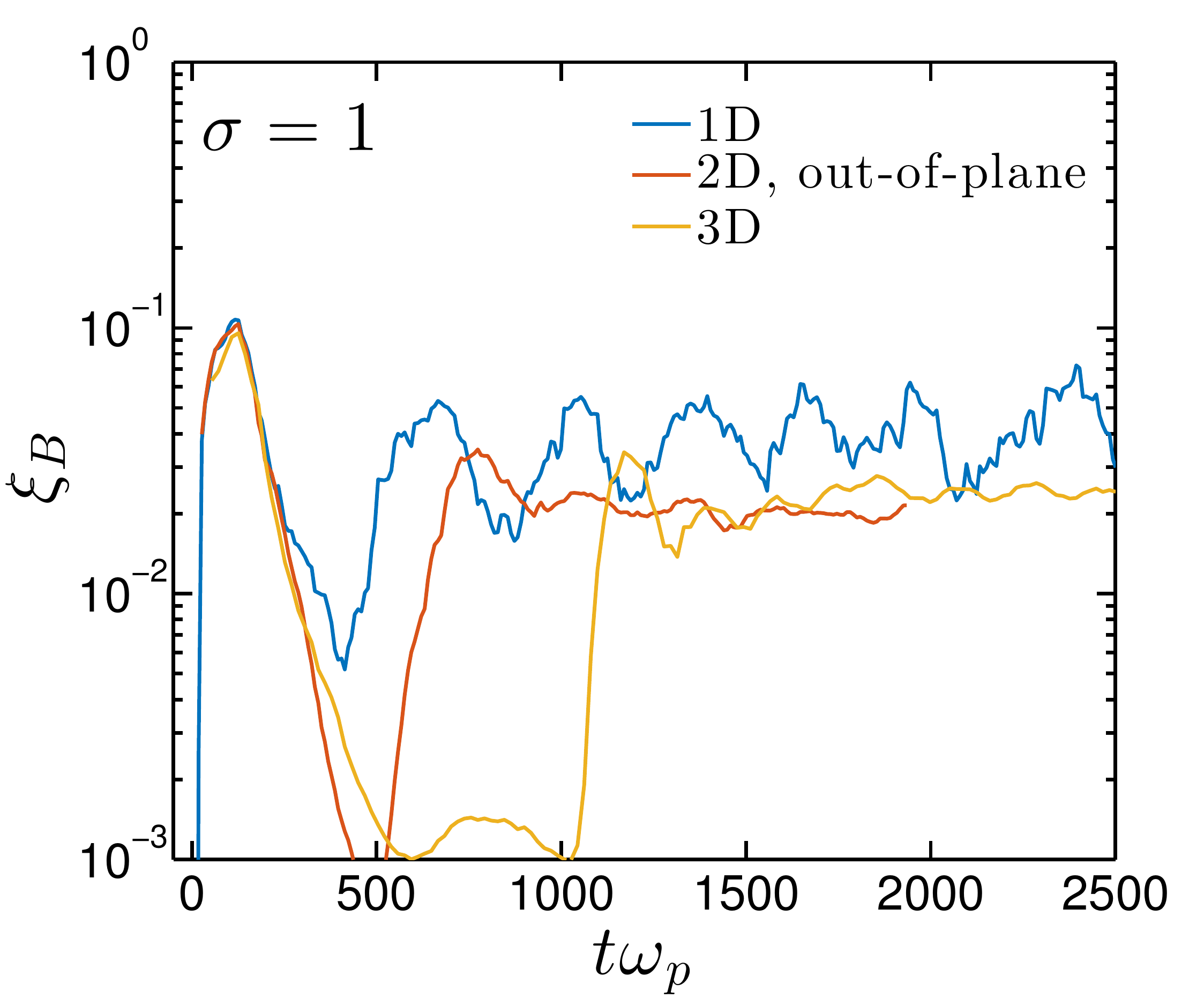}
		\caption{Comparison of the time-evolution of the escaping Poynting flux along the shock-normal direction in 1D, 2D and 3D simulations for $\sigma=1$. Blue solid line presents 1D, red for 2D (out-of-plane $B_{z,0}$), and orange for 3D. 1D results are the same as in Figure~\ref{fig:xiB_time_evolution}, as the Poynting flux is nearly equal to $\xi_B$ in 1D. See the main  text for details on numerical parameters in 2D and 3D simulations.}
		\label{fig:time_evol_sig1_1d2d3d}
	\end{center}
\end{figure}

\begin{figure}
	\begin{center}
		\includegraphics[width=0.48\textwidth,angle=0]{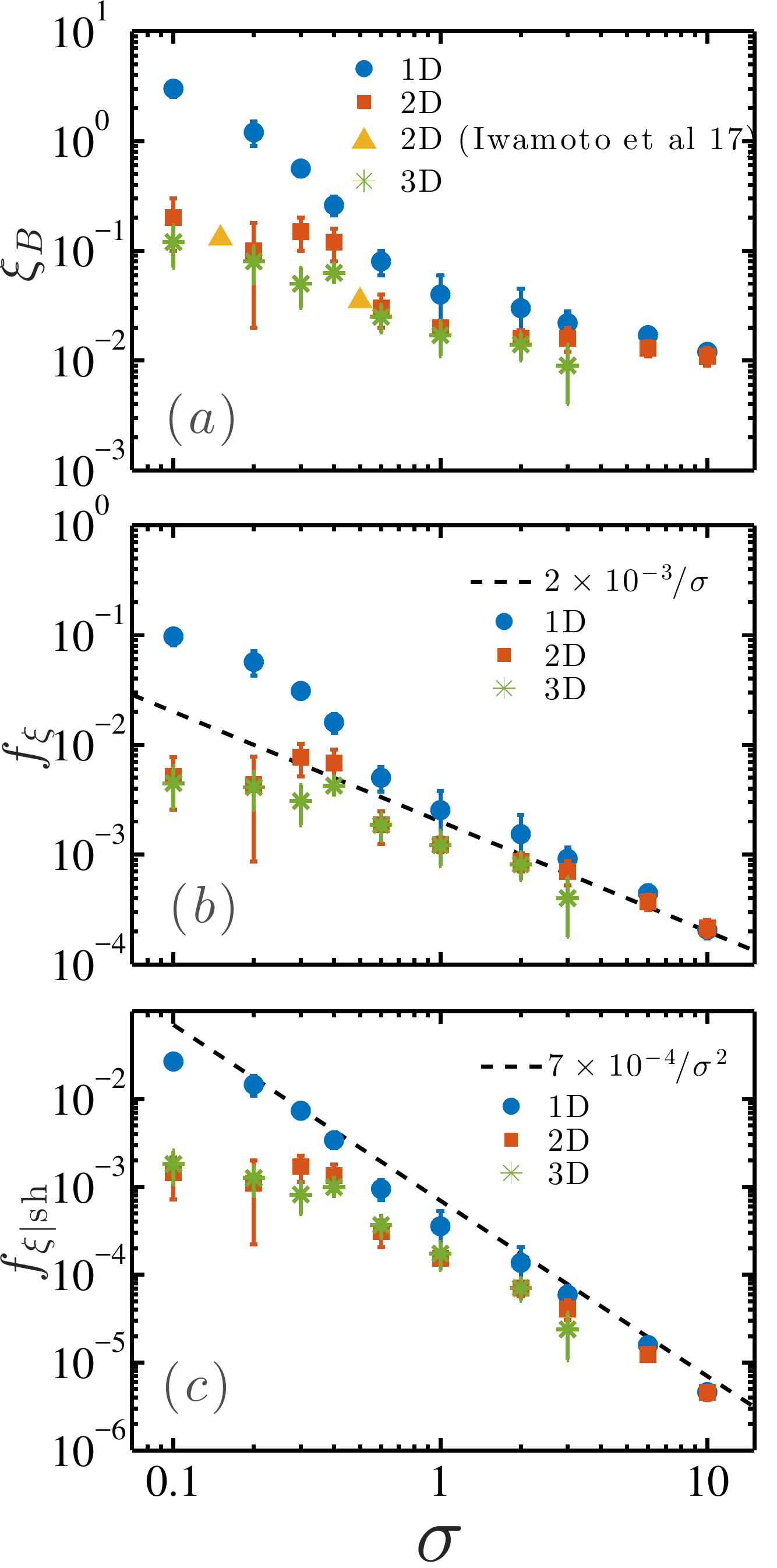}
		\caption{Dependence on the upstream magnetization $\sigma$ of the normalized Poynting flux of the precursor wave $\xi_B$ (panel a), of the energy fraction parameter measured in the simulation frame $f_\xi$ (panel b), and of the energy fraction measured in the shock front rest frame $f_\xi^\prime=f_{\rm \xi | sh}$ (panel c). The dashed line in panel~b follows the scaling $f_\xi = 2\times 10^{-3}/\sigma$, in the same way as in Figure~\ref{fig:xiB_sigma_3panel}. The dashed line in panel~c follows $f_{\xi | \rm sh} = 7\times 10^{-4}/\sigma^2$, identically to Figure~\ref{fig:SRF_energy_budget}.}
		\label{fig:3panel_prec_energy_1D2D3D}
	\end{center}
\end{figure}

In order to support our claim that the precursor wave energy does not significantly decrease due to multi-dimensional effects (in the $\sigma \geq 1$ regime of interest for this work), here we present a preliminary analysis of 2D and 3D simulations performed with Tristan-MP. \footnote{The code accuracy and stability in multi-dimensional simulations of relativistic shocks was assessed in several studies \citep{spitkovsky_05, spitkovsky_08b, sironi_spitkovsky_09, sironi_spitkovsky_11, sironi_13}.} We explore a range of $\sigma \in [0.1, 10]$ in 2D, and $\sigma \in [0.1, 3]$ in 3D. In 2D simulations we focus on the \textit{out-of-plane} configuration: the simulation plane is the \textit{xy} plane, the shock front propagates in the $x$-direction, and the upstream magnetic field is along the $z$-direction. We do not present any \textit{in-plane} 2D simulation results here because we find that 3D simulations are in excellent agreement with 2D out-of-plane results.

	In 2D simulations we keep all parameters the same as in 1D, except that the number of particles per cell per species is set to 8 (values between 2 and 32 have been tested with no significant differences). The transverse dimension of the simulation box is set to $14 \,c/\omega_{\rm p}$.We find that a transverse width of more than $2-3 \,c/\omega_{\rm p}$ is sufficient to capture multi-dimensional effects. In particular, the effects of wave filametation and self-focusing that lead to efficient  pre-heating of the upstream plasma in the longitudinal momentum are properly captured with a box width of a few skin depths.
	
	In 3D simulations we reduce the transverse dimension to $4\, c/\omega_{\rm p}$ (in both $y$ and $z$ directions). The spatial resolution in 3D runs is set to 25 cells per $c/\omega_{\rm p}$ (four times lower than in 1D and 2D) and the number of particles per cell per species is varied between 3 and 18 (again, with little differences). This was necessary to produce sufficiently long runs while still capturing the relevant physics. The effect of a lower spatial resolution was only apparent in the $\sigma=3$ run, since the spectrum extends to higher frequencies, which are not captured properly if the resolution is insufficient.
	
	Figure~\ref{fig:time_evol_sig1_1d2d3d} presents the time evolution of the precursor energy for $\sigma=1$ as measured in 1D (blue line), 2D (red line) and 3D (orange line) simulations. The direct comparison between 1D and multi-dimensional simulations shows that the asymptotic value of $\xi_B$ in 2D and 3D simulations is only a factor of two smaller than in 1D, while for $\sigma < 0.5$ --- as we will show below, and see also \citet{iwamoto_17} --- the energy of the wave decreases by a factor of about $3-10$ when going from 1D to 2D and 3D configurations. It also shows that the 2D and 3D energetics are in very good agreement. The only difference between 2D and 3D is that it takes more time in 3D to settle into the steady state (see the rise of the red line after $t\omega_{\rm p}=500$ and of the orange line after $t\omega_{\rm p}=1000$). So, we can confidently state that the decrease in precursor efficiency due to multi-dimensional effects is much less severe in the high-magnetization case $\sigma=1$ than for $\sigma <0.5$.
	
	 Let us note that in this appendix we have redefined the $\xi_B$ parameter. Here, $\xi_B$ corresponds to the normalized Poynting flux in the $x$-direction, $\xi_B=\langle \delta E_y \delta B_z - \delta E_z \delta B_y \rangle/B_0^2$. The average is done over the region between $5$ and $25 \,c/\omega_{\rm p}$ ahead of the shock front, for consistency with our 1D results, and over all the transverse directions ($y$  in 2D; $y$ and $z$  in 3D). In 1D we have systematically verified that $\langle \delta B_z^2 \rangle = \langle \delta E_y \delta B_z - \delta E_z \delta B_y \rangle=\langle \delta E_y \delta B_z \rangle$, but this equality is not obviously satisfied in multi-dimensional simulations with $\sigma \leq 0.6$. The choice of defining $\xi_B$ as the precursor Poynting flux is due to the fact that the most relevant measure of the electromagnetic energy output of the shock is the Poynting flux of the escaping wave in the shock-normal direction.
	 
  In Figure~\ref{fig:3panel_prec_energy_1D2D3D}, using a suite of 1D, 2D and 3D simulations, we show the dependence on $\sigma$ of the normalized Poynting flux of the precursor wave $\xi_B$ (panel a), of the energy fraction parameter as measured in the simulation frame $f_\xi$ (panel b), and of the energy fraction parameter as measured in the shock rest frame $f_\xi^\prime=f_{\rm \xi | sh}$ (panel c). The definition of the latter two is given in the main body of the article: Eq.~\ref{eq:fxi_DRF} and Eq.~\ref{eq:fxi_SRF}, respectively. Values from 1D, 2D and 3D simulations are plotted using blue circles, red squares and green stars, respectively. The results of 2D out-of-plane simulations of \citet{iwamoto_17} are plotted using orange triangles in panel (a). 
  The measurement of $\xi_B$ in 2D and 3D simulations was done by considering the asymptotic values in the time evolution  for each $\sigma$, as shown in Figure~\ref{fig:time_evol_sig1_1d2d3d} for the particular case of $\sigma=1$. Error bars quantify uncertainties due to temporal oscillations of the time-evolution curves. Knowing $\xi_B$ and measuring directly the shock front velocities from simulations, the values in panels (b) and (c) were produced using Eq.~\ref{eq:fxi_DRF} and Eq.~\ref{eq:fxi_SRF}, respectively.
  
Figure~\ref{fig:3panel_prec_energy_1D2D3D} shows that:
  \begin{itemize}
  	\item In 2D and 3D (red and green symbols), for $\sigma=0.1$ the Poynting flux of the precursor wave $\xi_B$, the energy fractions $f_\xi$ and $f_{\xi|\rm sh}$ are reduced by a factor of $\approx 10-20$ as compared to 1D (blue circles). This is in agreement with \citet{iwamoto_17}.
  	\item The suppression in efficiency becomes gradually smaller when $\sigma$ increases from $0.1$ to $3$. For $\sigma \gtrsim 1$, the difference between 1D and multi-dimensional results becomes negligible.
  	\item Values from 2D out-of-plane and 3D simulations are generally in very good agreement, except for $\sigma=0.3$ and $0.4$ (which we have called ``transition cases'' in the main body of the text).
  	\item If the precursor energy fraction is cast in the shock rest frame, panel (c) shows that  $f_{\xi}^\prime \simeq 10^{-3}$ for $\sigma \sim 0.1-0.4$, instead of $\sim 0.01$ in 1D. For $\sigma >1$, multi-dimensional simulations converge towards 1D values and follow the scaling $f_{\xi}^\prime \approx 5 \times 10^{-4} /\sigma^2$, only slightly lower than reported in the main text for 1D simulations only.
  \end{itemize}
Using 3D simulations we can address other aspects of the precursor physics, such as the importance of the O-mode ($\delta B_y$ component, since $\mathbf{\delta B \perp B_0}$ for this mode) versus X-mode ($\delta B_z$ component, since $\mathbf{\delta B \parallel B_0}$ for this mode) and beaming of the emitted precursor wave. By extracting systematically the values of $\langle \delta B_y^2 \rangle$ and $\langle \delta B_z^2 \rangle$ in 3D simulations, we find that the O-mode is subdominant for all magnetizations explored here, i.e., $\langle\delta B_y^2\rangle/\langle\delta B_z^2\rangle \sim 10^{-3}$. This implies that the precursor wave retains (at the $99\%$ level, or more) the linear polarization of the X-mode, with magnetic field of the wave lying in the same direction as the upstream background field.

Concerning the beaming of the precursor wave in 3D, we considered the components of the Poynting vector in different directions. We find that the Poynting flux along the $y$-direction (and $z$-direction) is largely subdominant as compared to the shock-normal direction. The ratio is $|\Pi_y|/\Pi_x \sim 5 \times 10^{-4}$ for any $\sigma \in [0.1, 3]$, where the Poynting vector of the wave is defined as $\mathbf{\Pi}=\mathbf{\delta E} \times \mathbf{\delta B}/B_0^2$. This shows that the emitted wave is strongly beamed in the shock-normal direction. For an external observer the beaming will be further enhanced by Lorentz transformation from the simulation frame to the observer frame (in the case of shocks in magnetar winds, from the post-shock frame to the pre-shock frame).

This preliminary analysis of multi-dimensional runs demonstrates that 1D simulations provide accurate numbers in the $\sigma \gg 1$ regime, in agreement with 2D out-of-plane and 3D simulations.

\end{document}